\documentstyle[12pt]{article}
\input psfig

\def\reference{\parskip 0pt\par\noindent\hangindent 0.5 truecm}

\begin{document}

\title{The Optical/Near-IR Colours of Red Quasars}

\author{Paul J Francis $^{1,2}$ \and
 Matthew T. Whiting $^{3}$ \and
Rachel L. Webster $^{3}$
}

% Date - leave this blank.
\date{}
\maketitle

{\center
$^1$ Research School of Astronomy and Astrophysics, Australian National
University, Canberra ACT 0200\\pfrancis@mso.anu.edu.au\\[3mm]
$^2$ Joint appointment with the Department of Physics and Theoretical
Physics, Faculty of Science\\[3mm]
$^3$ School of Physics, University of Melbourne, Parkville, VIC 3052
\\mwhiting,rwebster@physics.unimelb.edu.au\\
[3mm]
}

% Abstract
% Simply place your abstract between the \begin{abstract} and
% \end{abstract} commands.
%
\begin{abstract}
% Place the abstract here.

We present quasi-simultaneous multi-colour optical/near-IR
photometry for 157 radio selected quasars, forming an unbiassed
sub-sample of the Parkes Flat-Spectrum Sample. Data are also presented
for 12 optically selected QSOs, drawn from the Large Bright QSO Survey.

The spectral energy distributions of the radio- and optically-selected
sources are quite different. The optically selected QSOs are
all very similar: they have blue spectral energy distributions curving
downwards at shorter wavelengths. Roughly 90\% of the radio-selected quasars 
have roughly power-law spectral energy distributions, with slopes
ranging from $F_{\nu} \propto \nu^0$ to $F_{\nu} \propto \nu^{-2}$.
The remaining 10\% have spectral energy distributions showing
sharp peaks: these are radio galaxies and highly reddened
quasars.

Four radio sources were not detected down to magnitude limits of
$H \sim 19.6$. These are probably high redshift ($z >3$) galaxies or
quasars.

We show that the colours of our red quasars lie close to the
stellar locus in the optical: they will be hard to identify in
surveys such as the Sloan Digital Sky Survey. If near-IR photometry
is added, however, the red power-law sources can be clearly separated
from the stellar locus: IR surveys such as 2MASS should be capable
of finding these sources on the basis of their excess
flux in the $K$-band.

\end{abstract}

{\bf Keywords:}
% Place keywords here.  PASA uses the standard list of subject 
% headings adopted by The Astrophysical Journal and available from URL:
%   http://www.noao.edu/apj/keywords96.html
Quasars: general - Methods: observational
\bigskip

\section{INTRODUCTION}

It was long believed that quasars are blue. The 
optical/near-IR colours of optically selected QSOs are indeed
uniformly very blue (eg. Neugebauer et al. 1987, Francis 1996).
It was therefore a surprise when substantial numbers of extremely
red quasars were identified in radio-selected samples (eg. Rieke, Lebofsky
\& Wisniewski 1982, Ledden \& O'Dell 1983, Webster et al. 1995, 
Stickel, Rieke, K\"{u}hr 1996). The biggest sample of these objects
is that of Webster et al, who were studying a sample of radio-loud quasars 
with flat radio spectra: the Parkes Half-Jansky Flat-Spectrum survey,
a complete sample of 323 sources with fluxes at 2.7 GHz ($S_{\rm 2.7}$)
of greater than 0.5 Jy, and radio spectral indices $\alpha$ 
($S_{\nu} \propto \nu^{\alpha}$) with $\alpha > -0.5$ as measured 
between
2.7 and 5.0 GHz (Drinkwater et al. 1997). While some of these
Parkes sources had $B_J-K_n$ colours as blue as any optically
selected QSOs, most had redder $B_J-K_n$ colours, and some
were amongst the reddest objects on the sky.

Why should the Parkes sources be so red? A variety of theories
were proposed:

\begin{itemize}

\item The $B_J$ magnitudes of the Parkes sample were measured
many years before the $K_n$ magnitudes. Quasars with flat radio
spectra are known to be highly variable: this could thus introduce
a scatter into the $B_J-K_n$ colours, though it is hard to see why it
should introduce a systematic reddening.

\item Elliptical galaxies with redshifts $z>0.1$ have very red
$B_J-K_n$ colours, due to the redshifted 400nm break. If the
host galaxies make a significant contribution to the integrated
light from the Parkes sources, this could produce the red colours.
Masci, Webster \& Francis (1998), however, used spectra to show
that this effect was only significant for 
$\sim 10$\% of the sample.

\item The $B_J$ magnitudes were derived from COSMOS scans of 
UK Schmidt plates, and are subject to substantial systematic
errors, which could introduce scatter into the $B_J-K_n$ colours
(O'Brian, Webster \& Francis, in preparation), though this too should
not introduce a systematic reddening.

\item Parkes quasars could have the same intrinsic colours as
optically selected QSOs, but be reddened by dust somewhere
along the line of sight (Webster et al. 1995).

\item Flat-radio-spectrum quasars are thought to have relativistic
jets: if the synchrotron emission from these jets has a very
red spectrum and extended into the near-IR, it could account for
the red colours (Serjeant \& Rawlings 1996).

\end{itemize}

In this paper, we test Webster et al's results by obtaining much better
photometry of a large sub-set of the Parkes sources. To minimise
the effects of variability, all our photometry for a given source
was obtained within a period of at most six days. All the data were
obtained from photometrically calibrated images, and rather than relying
on only two bands ($B_J$ and $K_n$), we obtained photometry in every
band from $B$ to $K_n$.

\begin{figure}
\psfig{file=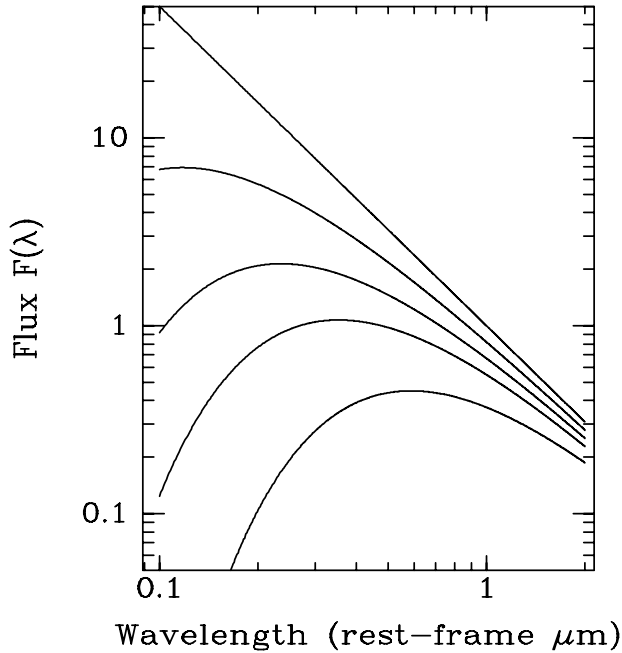,height=100mm}
\caption{Continuum shapes of dust affected quasars. The extinction
$E(B-V)$ increases downwards: values are 0, 0.1, 0.2,
0.3, 0.4. Note the characteristic `n' shape.\label{dustcont}}

\end{figure}

In principle, multi-colour photometry should enable us to discriminate
between the dust and synchrotron models. If quasars have intrinsically
blue power-law continua (eg. $F_{\nu} \propto \nu^{-0.3}$, Francis 1996), 
reddened by a foreground dust screen with an extinction $E(B-V)$ between
the $B$ and $V$ bands (in magnitudes) and an optical depth inversely 
proportional to wavelength, then the observed continuum slope 
will be

\begin{equation}
\label{dustlaw}
F_{\lambda} \propto  e^{- 2E(B-V)/\lambda} \lambda^{-1.7}, \label{eqn}
\end{equation}
where $\lambda$ is the wavelength in $\mu$m. This is
plotted in Fig~\ref{dustcont}. Note the very characteristic 
`n' shape, as the dust absorption increases exponentially into the
blue.

\begin{figure}
 \psfig{file=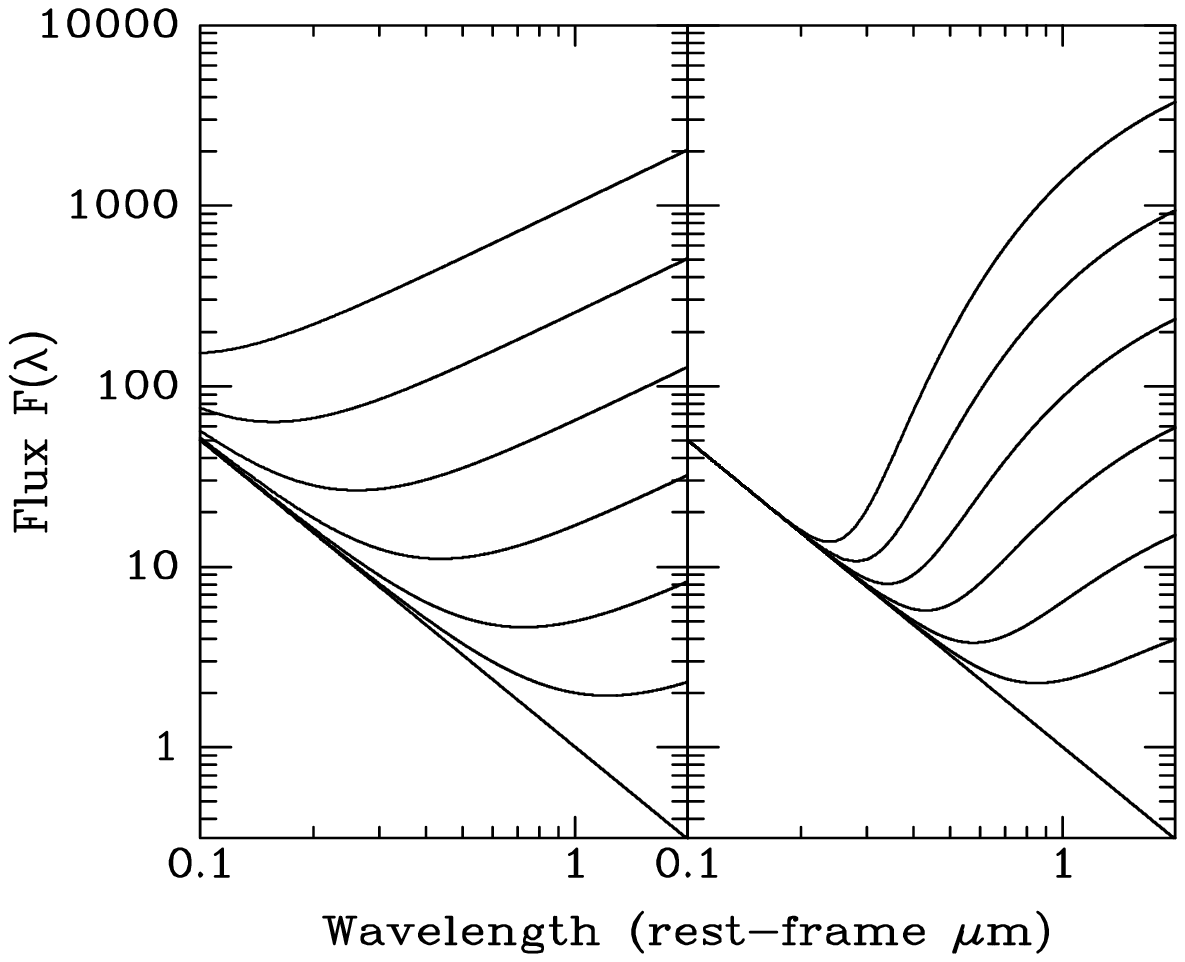,height=100mm}
\caption{Continuum shapes of quasars with an additional red emission
component. To show some of the possibilities, two different arbitrary
functional
forms have been chosen for the red component: a power-law (left) and
an exponential (right). The strength of this red component increases
upwards. Note the characteristic `u' shape.\label{redcont} The plausibility 
of synchrotron models is discussed in Whiting, Webster \& Francis (2000).}

\end{figure}

If, alternatively, the redness is caused by the addition of some red
synchrotron emission component to the underlying blue continuum, continuum
shapes will have a characteristic `u' shape, dominated by the underlying
blue flux at short wavelengths but by the new synchrotron component at
longer wavelengths (Fig~\ref{redcont}).

If radio-quiet red quasars exist, they cannot be selected by conventional
optical surveys. We show that by combining optical and near-IR data, it
should be possible to select any radio-quiet sources with the colours of 
most of our radio-selected red quasars.

This paper describes the observations, presents the data,
includes some simple phenomenological analyses of the results, and
discusses the colour selection of red quasars in the optical and
near-IR.
We defer the detailed modelling of the data to another paper:
Whiting, Webster \& Francis (2000).

\section{OBSERVATIONS}

We obtained quasi-simultaneous B, V, R, I, J, H and $K_n$ photometry of 
a subset of the Parkes sample. Observations were taken during 26 nights 
in 1997 (Table~1) at Siding Spring Observatory. Optical images were obtained
with either the 1 m telescope, or with the imager on the 2.3 m telescope.
Near-IR images were obtained with the CASPIR 256$\times$256
InSb array camera (McGregor et al. 1994) on the 2.3 m telescope. 157
Parkes sources were observed in some or all of the bands, as well
as a small control sample of 12 optically selected QSOs randomly
selected from the Large Bright QSO survey (LBQS, Morris et al. 1991);
an optical QSO survey well matched in size and redshift distribution
to the Parkes sample. To minimise the
effects of variability, all the observations of an individual
source were made within, at most, a six day period (Table~2). Flat
spectrum quasars typically vary by 10\% or less on these timescales,
though very occasional greater variations are seen, typically in BL Lac
objects (eg. Wagner et al. 1990, Heidt \& Wagner 1996). Only
data taken in photometric conditions were used: seeing was typically
1--2''. 

Bright objects were typically observed for $\sim$ five minutes in
each band. Fainter objects were observed for up to two hours in
our most sensitive bands ($R$, $I$ and $H$). If they were seen in
these bands, we observed them in progressively bluer bands as time
allowed.  Four sources were not detected in any band:
PKS~1535$+$004, PKS~1601$-$222, PKS~1649$-$062 and PKS~2047$+$098.

About five standard stars, spanning a range of colours, were
observed each night: in the optical,
the Graham E regions (Graham 1982) were used, while in the near-IR,
photometric calibration was obtained using the IRIS standard stars, 
which have magnitudes on the Carter SAAO system (Carter \& Meadows 1995).
Within individual nights, the scatter in photometric zero points
(without using colour corrections) was $< 3$\% rms, so all the standards
in a given band were simply averaged to give the final calibration.

\begin{table}
\begin{center}
\caption{Observing Log}
\begin{tabular}{lll}
\hline 
Night Code & Date & Telescope/Instrument \\
\hline \\
A & April 12, 1997 & 1m \\
B & April 13, 1997 & 1m \\
C & April 14, 1997 & 2.3m Imager \\
D & April 15, 1997 & 2.3m Imager \\
E & April 16, 1997 & 2.3m Caspir \\
F & April 17, 1997 & 2.3m Caspir \\
I &  July 12, 1997 & 1m \\
J &  July 13, 1997 & 1m \\
K &  July 14, 1997 & 1m \\
L &  July 13, 1997 & 2.3m Imager \\
M &  July 14, 1997 & 2.3m Imager \\
N &  July 15, 1997 & 2.3m Caspir \\
O &  July 16, 1997 & 2.3m Caspir \\
P &  July 17, 1997 & 2.3m Caspir \\
Q &  July 18, 1997 & 2.3m Caspir \\
R &  July 19, 1997 & 2.3m Caspir \\
S &  July 20, 1997 & 2.3m Caspir \\
T &   Sept 7, 1997 & 1m \\
U &   Sept 8, 1997 & 1m \\
V &   Sept 9, 1997 & 2.3m Imager \\
W &  Sept 10, 1997 & 2.3m Imager \\
X &  Sept 11, 1997 & 2.3m Caspir \\
Y &  Sept 12, 1997 & 2.3m Caspir \\
Z &  Sept 13, 1997 & 2.3m Caspir \\ \hline
\end{tabular}
\end{center}
\end{table}

All 98 Parkes sources lying in the RA. ranges 00:36 -- 00:57,
01:53 -- 02:40 and 14:50 -- 22:52 (B1950) were observed in both the
optical and the IR: these
should thus form an unbiassed, complete sub-sample of the whole Parkes
Half-Jansky sample. The remaining 59 sources were selected for
observation mainly on the basis of prevailing weather conditions,
and so should also form a reasonably unbiassed sub-sample. No
selection was made against radio galaxies: sources with resolved
optical or near-IR images (as classified by the COSMOS plate 
measuring machine from UK Schmidt plates, and checked by visual
inspection of our images) are listed in Table~2. Where appropriate,
they are excluded from the following analysis.

Optical images were bias- and overscan-subtracted, and then
flat fielded using twilight sky flats. For the fainter sources,
multiply dithered 300- or 600-second exposures were taken: these 
were combined using inverse variance weighting. The infrared
exposures were made up of multiple dithered 60 second images,
each made up of two averaged 30 sec exposures in $J$, six averaged
10 sec exposures in $H$ and twelve averaged 5 sec exposures in
$K_n$. These were bias- and dark-subtracted, and then corrected for
the non-linearity of the CASPIR detector using a simple quadratic
correction term (derived from plots of median counts against exposure 
time obtained from dome flats). Known bad pixels were replaced by
the interpolated flux from neighbouring pixels.
Flat fields were obtained by taking
exposures of the dome with lamps on and off, and subtracting one from the
other: this removes the contribution from telescope emission, and
substantially improves the photometric accuracy attainable. Individual
images were sky subtracted, using a median of the 10 images taken 
nearest in time. The dithered images were then aligned and combined,
using the median to remove residual errors.

The radio sources were identified from the radio positions by using
astrometry from nearby stars, bootstrapped from positions in the
COSMOS/UKST and APM/POSS sky catalogues, maintained on-line at the
Anglo-Australian Observatory. Magnitudes were then measured using
circular apertures, with the sky level determined from the median
flux in an annulus around the sky aperture. For unresolved sources, 
the photometric apertures were set by the seeing: typical aperture
radii were $\sim 5$''. For resolved sources (mostly low redshift
radio galaxies) larger circular apertures were used, centred on
the galactic nucleus. These larger aperture radii are listed in 
the footnotes to Table~2. Standard stars were measured with
similar aperture sizes.

Quoted errors are the sum (in quadrature) of random errors
and an assumed 5\% error in the photometric zero points. Random errors were
determined by measuring the rms (root-mean-squared) pixel-to-pixel 
variation in sky regions, and scaling to the aperture size used. 
This will be accurate for fainter (sky or read-noise limited)
sources, but will underestimate random errors for the brightest
few sources. The photometric zero point errors were estimated from
the scatter in zero points between different
standard star measurements in an individual night: typical
rms scatters are $< 3$\%, so we adopted a conservative value of
5\% as our zero point error.

For modelling and plotting purposes, we converted the magnitudes
into fluxes. We assumed fluxes for zero magnitude objects as
listed in Table~\ref{zpoint}. In the optical, our filter set approximate the
Johnson \& Cousins system, and were calibrated using the Graham
standards (also approximating Johnson \& Cousins). The zero magnitude 
star fluxes for this system were taken from Bessell
Castelli \& Plez (1998). In the infrared, our observations used
the CASPIR filter set calibrated by the IRIS standards. Zero
magnitude fluxes were calculated by P. McGregor, assuming that Vega is
well represented in the near-IR by a black body of temperature
11200 K, and normalisation $F_{\lambda}{\rm (555 nm)} = 3.44 \times
10^{-12} {\rm W\ cm}^{-2}\mu m^{-1}$ (Bersanelli, Bouchet \&
Falomo 1991). These normalisations agree closely with those quoted for
UKIRT near-IR standards (MacKenty et al. 1997).
Our observations were made with the $K_n$ filter, but
were calibrated using the quoted $K$ magnitudes of the IRIS
standards without applying a colour correction term, and should thus
be normalised to a $K$-band zero point.

\setcounter{table}{2}
\begin{table}
\begin{center}
\caption{Assumed Fluxes of a Zero Magnitude Star\label{zpoint}}
\begin{tabular}{lcc}
\hline 
Filter & Mean & Flux of Zero Magnitude \\
 ~ ~   & Wavelength ($\mu$m) & Star ($F_{\lambda}$, 
${\rm W\ m}^{-2}{\rm nm}^{-1}$) \\
\hline \\
B & 0.440 & $6.32 \times 10^{-11}$ \\
V & 0.550 & $3.64 \times 10^{-11}$ \\
R & 0.700 & $2.18 \times 10^{-11}$ \\
I & 0.880 & $1.13 \times 10^{-11}$ \\
J & 1.239 & $3.11 \times 10^{-12}$ \\
H & 1.649 & $1.15 \times 10^{-12}$ \\
K & 2.132 & $4.10 \times 10^{-13}$ \\ \hline
\end{tabular}
\end{center}
\end{table}

\section{Results and Discussion}

\subsection{The Colour Distribution}

The results are listed in Table~4. Quoted errors are $1 \sigma$;
upper limits are $3 \sigma$. 

\begin{figure}
\begin{center}
\psfig{file=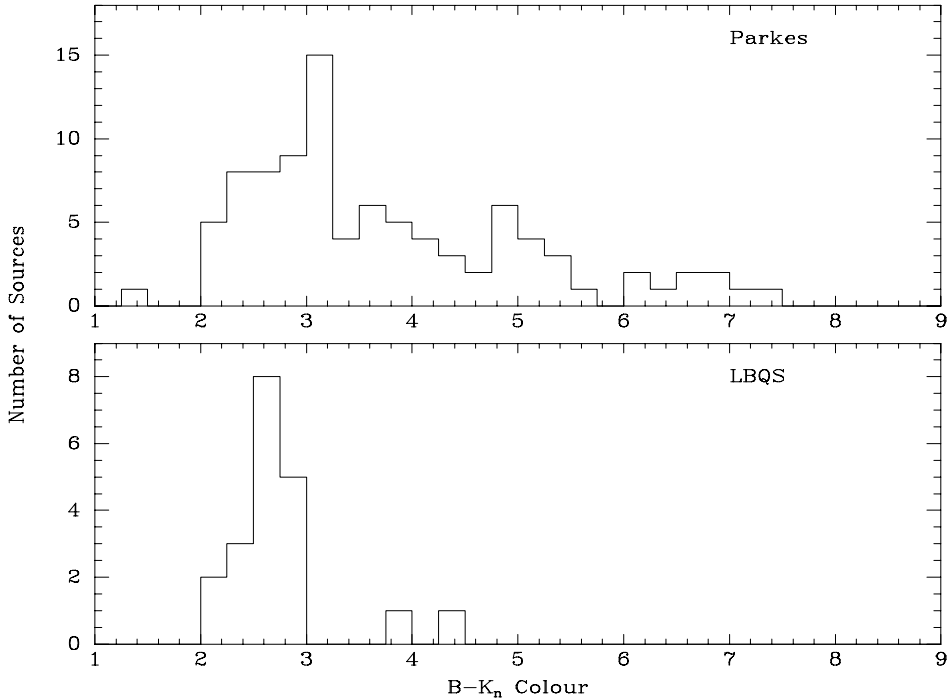,height=10cm}
\caption{The distribution of $B-K_n$ colours for the
Parkes sample (top panel), and the optically selected
LBQS sample (bottom panel). Sources with spatially
extended images (radio galaxies) have been excluded,
as have sources with redshift $z>3$ (as Ly$\alpha$
forest depressed the $B$-band flux). Only Parkes
sources within the complete sub-sample have been
used. The LBQS data from this paper have been
supplemented by data from Francis (1996).\label{bkfig}}
\end{center}
\end{figure}

Our data confirm the basic result of Webster et al: the Parkes
quasars have very different $B-K$ colours from optically selected
QSOs (Fig~\ref{bkfig}). The difference is significant: a 
Kolmogorov-Smirnov test shows that the the probability of getting
two samples this different from the same parent population is only
$9.1 \times 10^{-5}$. The bluest Parkes sources have colours very
similar to those of optically selected QSOs, but the distribution
of colours extends much further into the red. 

\begin{figure}
\begin{center}
\psfig{file=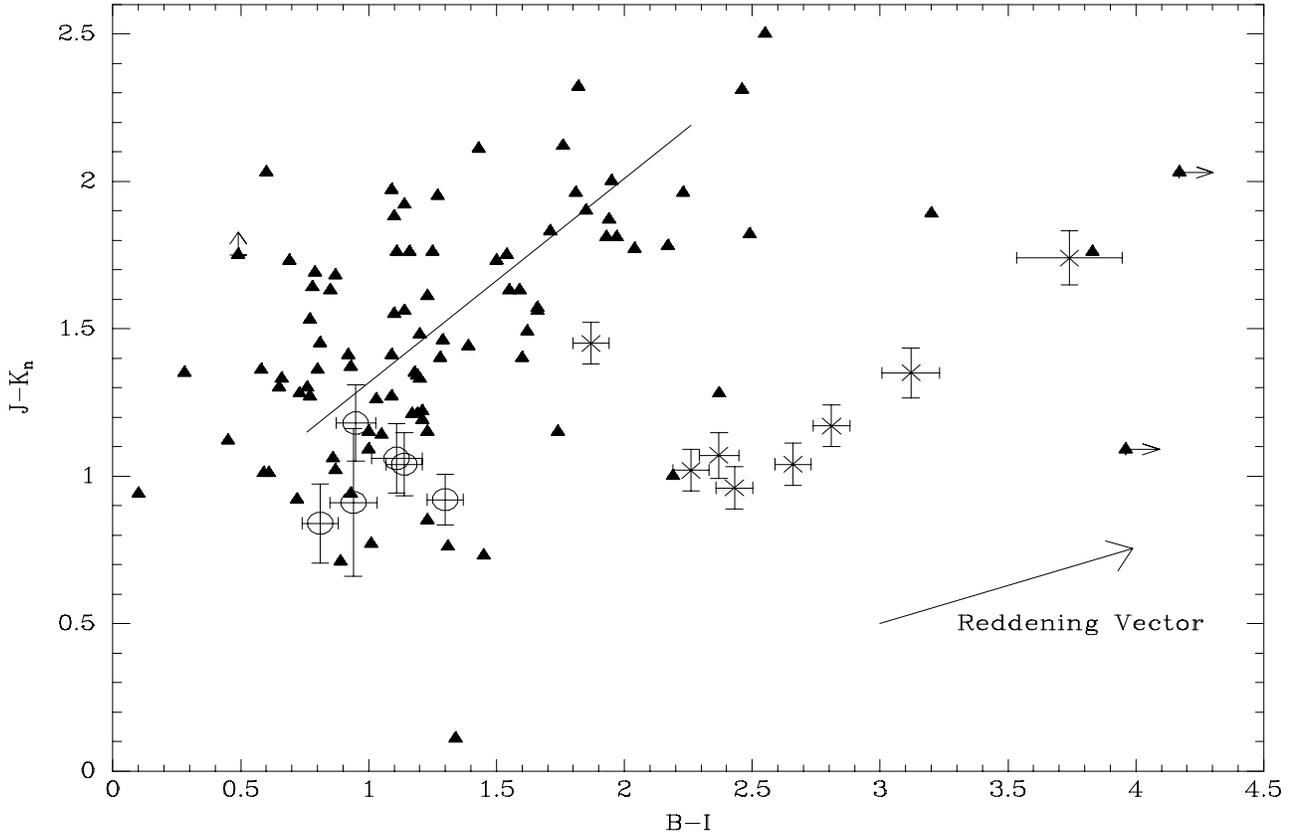,height=12cm}
\caption{The optical and infra-red colours of the complete
sub-set of the Parkes sample (triangles and crosses), compared
with a small sample of optically selected LBQS QSOs (circles).
Solid triangles denote unresolved sources: crosses are galaxies.
The solid line shows where a pure power-law continuum slope
would lie: it runs from $F_{\nu} \propto \nu^{0}$ on the left end, to
$F_{\nu} \propto \nu^{-2}$ on the right end. Error bars are not
shown for the unresolved Parkes sources, but are comparable to
those of the optically selected QSOs. The reddening vector is for
an extinction $E(B-V)=0.2$, a redshift of one, and dust extinction as
in equation~\ref{eqn}. The direction of the reddening vector is independent 
of redshift.
\label{colours}}
\end{center}
\end{figure}

\subsection{The `Main Sequence'}

Are the Parkes sources uniformly red everywhere between
$B$ and $K_n$? In Fig~\ref{colours} we plot a measure of the
optical colour ($B-I$) against a measure of the near-IR colour
($J-K_n$) for the complete sub-sample. Objects whose continuum
shape approximates a featureless power-law all the way from
$B$ to $K_n$ should lie close to the solid line in this plot.

\begin{figure}
\begin{center}
\psfig{file=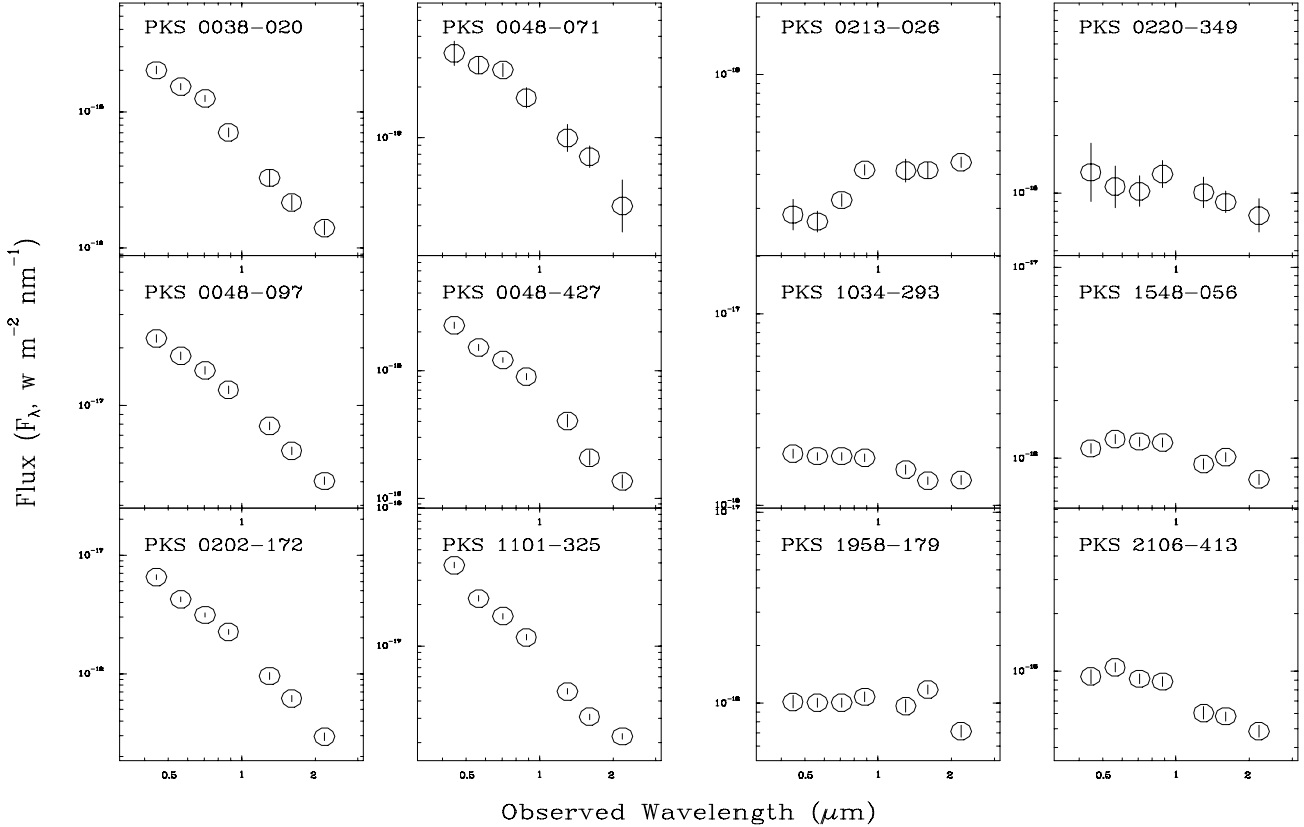,height=12cm}
\caption{Spectral energy distributions of representative 
Parkes quasars from the blue (left six plots) and red (right 
six plots) ends of the `main sequence', as defined in the text.
Sources on the left have $J-K_n < 1.5$ and $B-I < 1.5$; sources
on the right have $J-K_n > 1.8$ and $3 > B-I > 1.8$.
\label{mainseq}}
\end{center}
\end{figure}

$\sim 90$\% of all the Parkes sources do indeed lie 
close to the power-law line in Fig~\ref{colours}. These sources 
form a `main sequence'
of quasar colours, stretching from blue objects with
$F_{\nu} \propto \nu^{\sim 0}$ to red objects with
$F_{\nu} \propto \nu^{\sim -2}$. Examples of quasars from
both ends of this `main sequence' are shown in Fig~\ref{mainseq}.
Note that these quasars can lie on either side of the
power-law line: ie. they can have both `n'- and `u'-shaped
continuum spectra. The majority, however, lie above the line,
consistent with slightly `u'-shaped spectra (redder in the near-IR
than in the optical). This supports the synchrotron model
for these sources. We defer discussion of this point to the
detailed synchrotron modelling of the companion paper Whiting et al.

\subsection{Optically Selected QSOs}

\begin{figure}
\begin{center}
\psfig{file=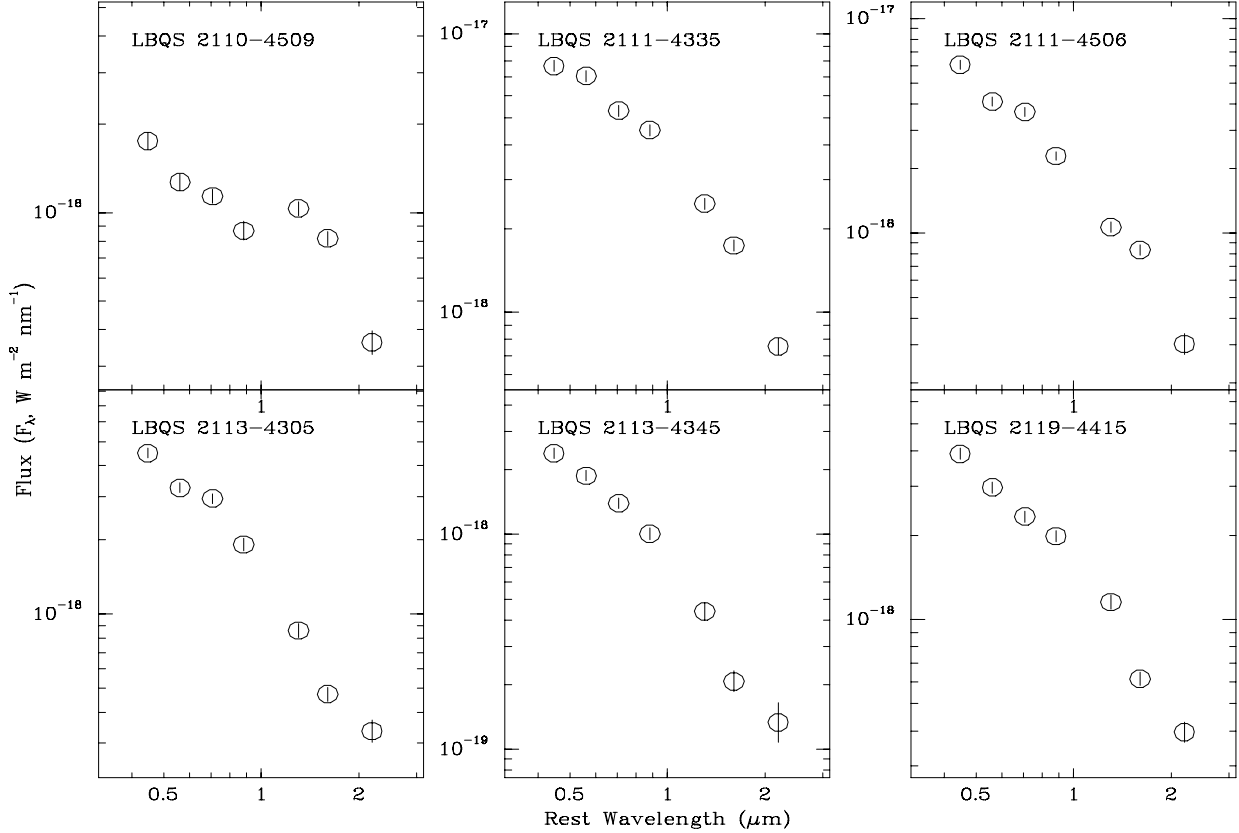,height=12cm}
\caption{Spectral energy distributions of all six optically
selected QSOs with complete photometric data.
\label{lbqs}}
\end{center}
\end{figure}

As Fig~\ref{colours} shows, the optically selected QSOs all have
very similar colours, and lie at the blue end of the `main
sequence'. They lie systematically below the power-law line,
however, indicating that they have `n' shaped spectra: ie. they are
redder in the optical than in the near-IR. This can be seen
in their spectra energy distributions, shown in Fig~\ref{lbqs}.

This spectral curvature matches the predictions of the dust
model. Wills, Netzer \& Wills (1985), however, suggested that it
may be partially due to blended Fe~II and Balmer-line emission, though
Francis et al. (1991) argued that this curvature is too large to be 
plausibly explained by emission-line contributions.

The position of the optically selected QSOs at the blue end of the
`main sequence' would be expected if the cause of redness in the
Parkes quasars is the addition of a red synchrotron component to
an underlying blue continuum which is identical to that in radio-quiet
QSOs (Whiting et al.).

\subsection{Galaxies and Extremely Red Objects}

\begin{figure}
\begin{center}
\psfig{file=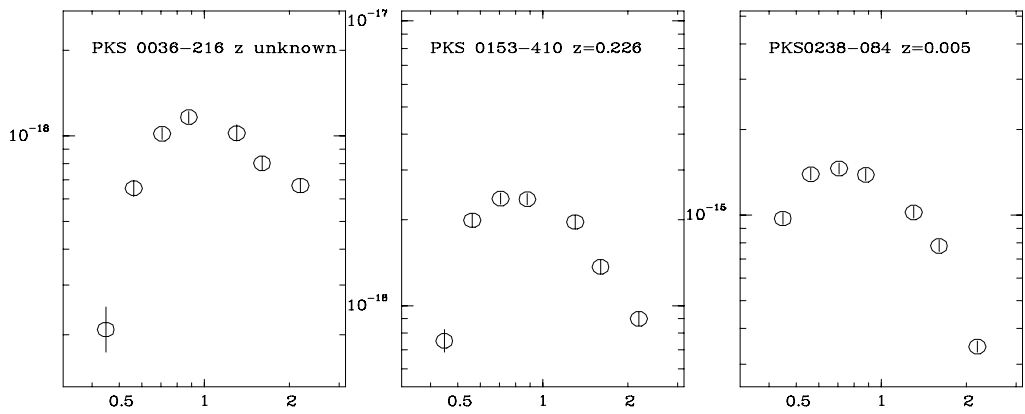,height=10cm}
\caption{The spectral energy distributions of three representative
galaxies from the Parkes sample.
\label{gals}}
\end{center}
\end{figure}

The colours of the spatially extended sources in the Parkes sample
are sharply peaked in the red, as would be expected from
moderate redshift galaxies (Fig~\ref{gals}). They therefore lie far
below the `main sequence' in Fig~\ref{colours}, the one exception
being PKS~1514$-$241, which is a galaxy at z=0.049 with a
BL Lac nucleus, which is presumably diluting the galaxy colours.
Higher redshift galaxies lie further to the right on this plot,
as would be expected due to the 400 nm break reducing the $B$-band
flux.

\begin{figure}
\begin{center}
\psfig{file=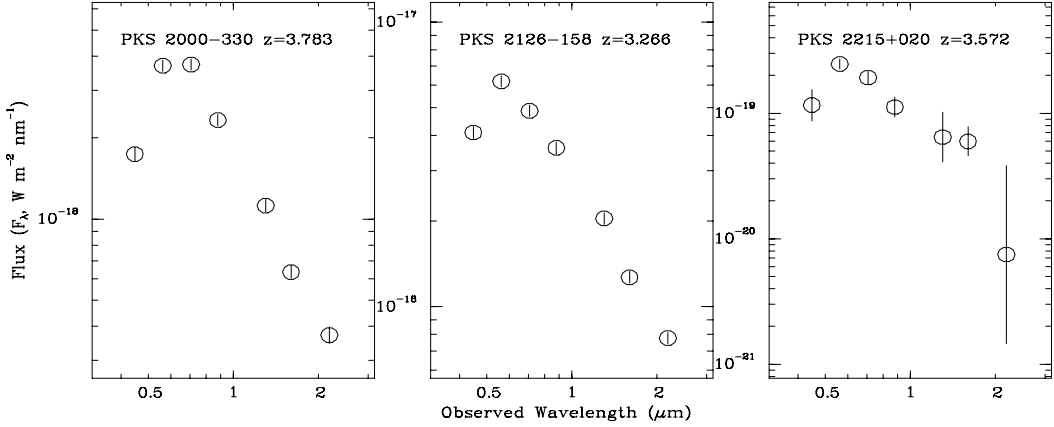,height=10cm}
\caption{The spectral energy distributions of three representative
Parkes quasars with redshifts $z>3$, showing the dip in the $B$-band
caused by Ly$\alpha$ forest absorption.
}
\end{center}
\label{highz}
\end{figure}

\begin{figure}
\begin{center}
\psfig{file=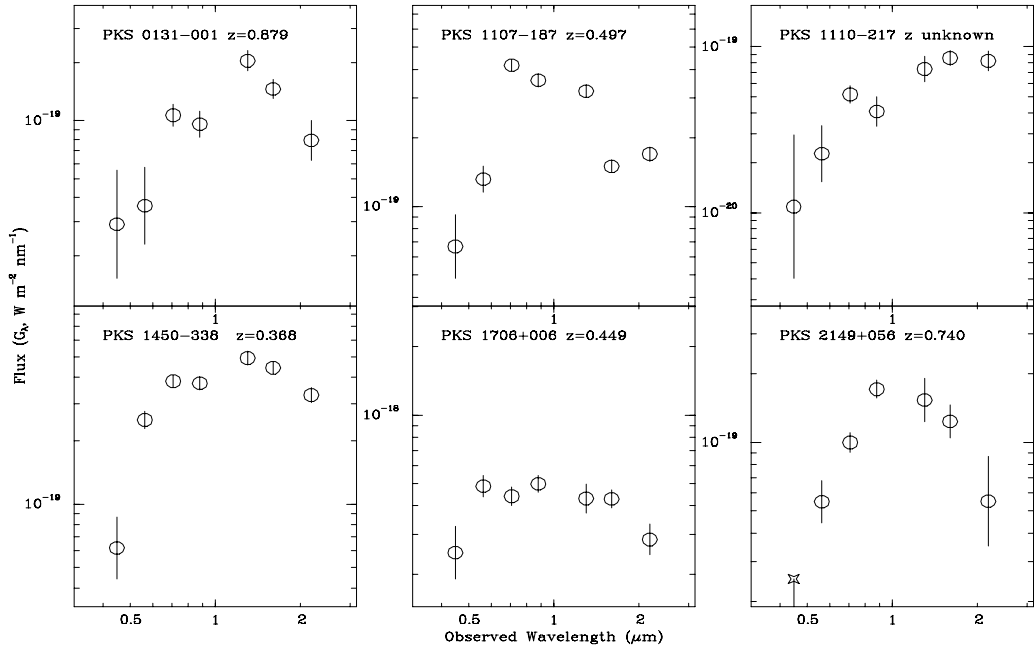,height=10cm}
\caption{The spectral energy distributions of the six Parkes sources
with $B-I>3$. The data for PKS~1706$+$006 have been adjusted for
galactic dust extinction of $E(B-V) = 0.23$ (Schlegel, Finkbeiner
\& Davis 1998), assuming a dust extinction law as described in the text.
\label{red}}
\end{center}
\end{figure}

\begin{figure}
\begin{center}
\psfig{file=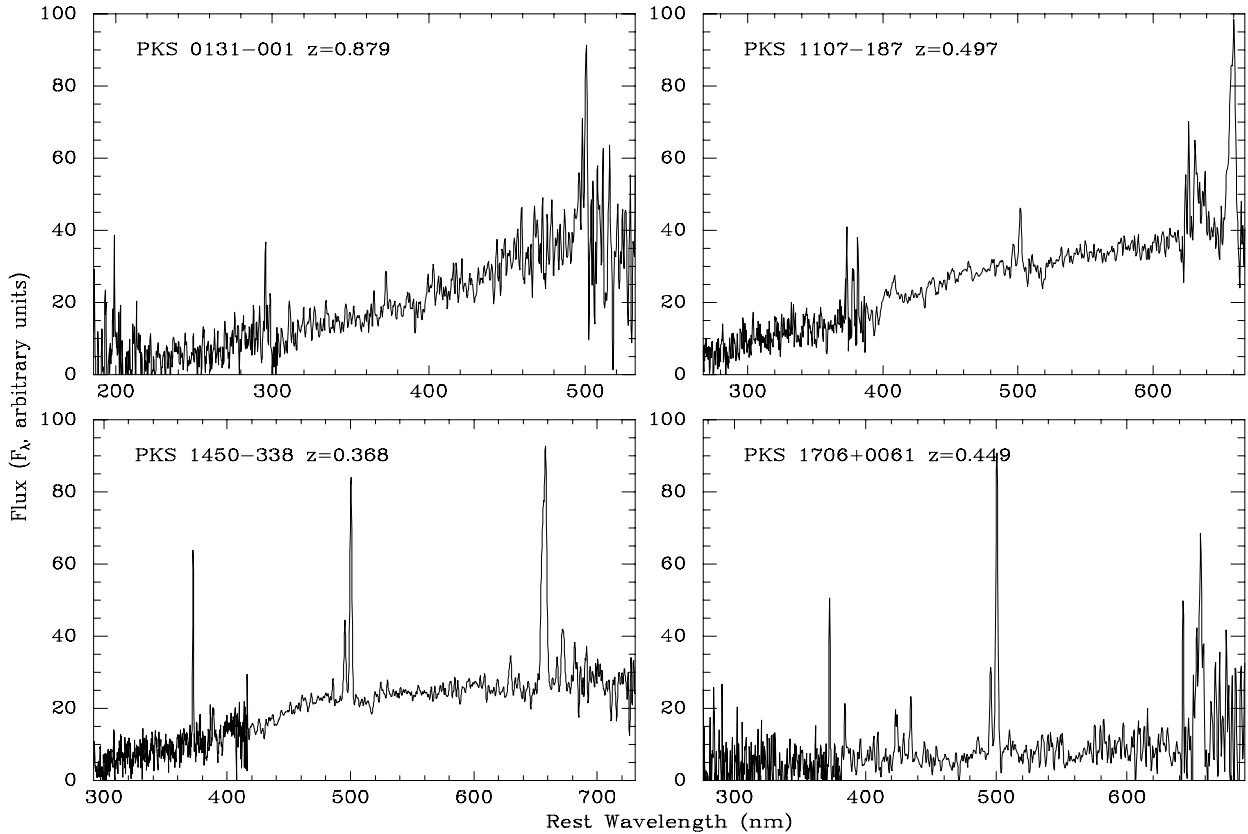,height=12cm}
\caption{Optical spectra of four extremely red Parkes sources.
With the exception of PKS~0131$-$001, the spectra show features
both of galaxy light (the 400 nm break and narrow [O~II] 372.7 nm
and [O~III] 495.9/500.7 lines) and of dust-reddened quasar light
(a red continuum at long wavelengths, broad H$\alpha$ 656.3 nm line
emission, and the notable weakness of the broad H$\beta$ 486.1 nm line
with respect to H$\alpha$).
\label{specplot}}
\end{center}
\end{figure}

What are the other, red, highly `n'-shaped objects lying far below the
`main sequence' which are not spatially resolved? A few are high redshift
QSOs, in which the $B$-band flux has been reduced by Ly$\alpha$ forest
absorption (Fig~\ref{highz}). The reddest objects, however, with $B-I>3$
(Fig~\ref{red}), do not lie at high redshifts. We have obtained
spectra of four of these very red objects (Francis et al. 2000,
in preparation). Three show hybrid spectra: they look like galaxies
at short wavelengths, but at longer wavelengths a red power-law
continuum component is seen, along with broad emission-lines. The
ratios of H$\alpha$ to H$\beta$ are around 20: far above those seen in
normal AGN ($\sim 5$) and evidence of substantial reddening 
(Fig~\ref{specplot}). Note that these hybrid objects all have radio
spectra indices near the steep spectrum cut-off of our sample, as
do the galaxies in the sample.

The reddest objects are thus a heterogeneous group: some are high
redshift quasars, some are galaxies, and some are heavily dust-reddened
quasars.

\subsection{Unidentified Objects}

Four Parkes sources were not detected in any band. After correction
for galactic foreground absorption (Schlegel et al.), our non-detections
impose 3$\sigma$ upper limits of $H> 19.61$ for PKS~1532$+$004,
$H > 19.76$ \& $K > 19.29$ for PKS~1601$-$222, $H>17.22$ and $K>16.61$
for PKS~1649$-$062 (which is subjected to substantial galactic reddening)
and $H > 19.82$ for PKS~2047$+$098.

If unified schemes for radio-loud AGN are correct, the host galaxies
of our flat-radio-spectrum sources should be very similar to those of
steep-radio-spectrum radio galaxies. This enables us to place
a lower-limit on the redshift of these unidentified sources: even if
their AGN light is completely obscured, we should still see the host
galaxy, which should lie on the $K$-band Hubble diagram for
radio galaxies (eg. McCarthy 1992). To be undetected at our
magnitude limits, therefore, all these sources must lie above
redshift 1, and apart from PKS~1649$-$062, probably lie
above redshift 3.

\subsection{Anomalous Objects}

\begin{figure}
\begin{center}
\psfig{file=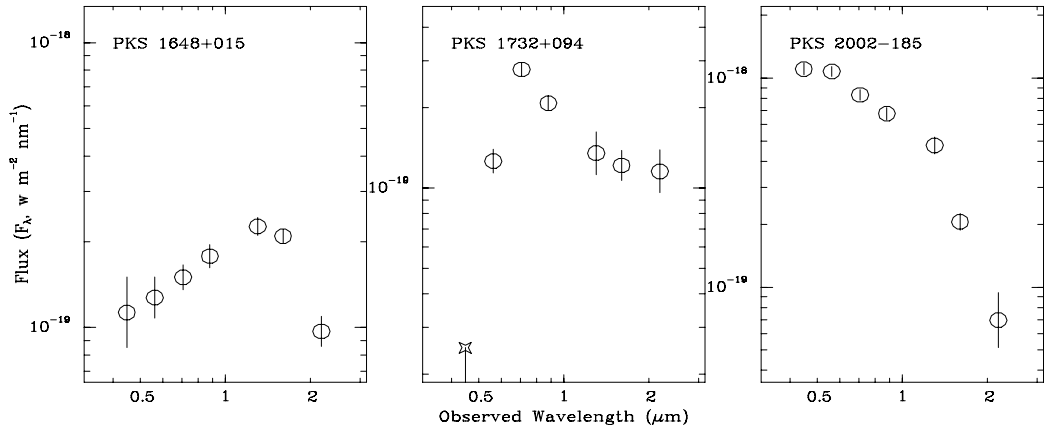,height=10cm}
\caption{Spectral energy distributions of three anomalous Parkes
sources.
\label{weirdo}}
\end{center}
\end{figure}

Three sources have colours that do not fit any of these categories
(Fig~\ref{weirdo}). We discuss these in turn.

PKS~1648$+$015 shows a smooth optical power-law
rising into the red, until at around $1.4 \mu$m, the flux abruptly
decreases. As all the IR data points were taken within minutes of
each other in good weather conditions, we believe that this near-IR
turn-over is real. We obtained a somewhat noisy optical spectrum of
this source (Drinkwater et al.) which shows a featureless, very red
power-law, in excellent agreement with the photometry. We cannot
explain this source.

PKS~1732$+$094 is blue longwards of around $0.6 \mu$m, but drops
dramatically at shorter wavelengths. Our spectrum of this source 
(Drinkwater et al.) is too poor to be of any use. We hypothesise
that this may be a very high redshift $z>4$ quasar, and that the
drop in the blue is due to Ly$\alpha$ absorption.

PKS~2002$-$185 has optical colours typical of the bluest Parkes
sources, but in the near-IR is bluer still: far bluer than any
other source at these wavelengths. An optical spectrum, covering
a very restricted wavelength range (Wilkes et al. 1983) shows
only a single broad emission-line: on the assumption that this is Mg~II
(279.8 nm) a redshift of 0.859 is determined.

\section{Multicolour Selection of Red Quasars}

Could there be a population of radio-quiet QSOs with the same
colours as our radio-loud red quasars? Webster et al. showed that it
is virtually impossible to find such QSOs in any sample with a blue
optical magnitude limit. In this section we ask whether red QSOs
could be identified by colour selection in the red optical and near-IR.

\begin{figure}
\begin{center}
\psfig{file=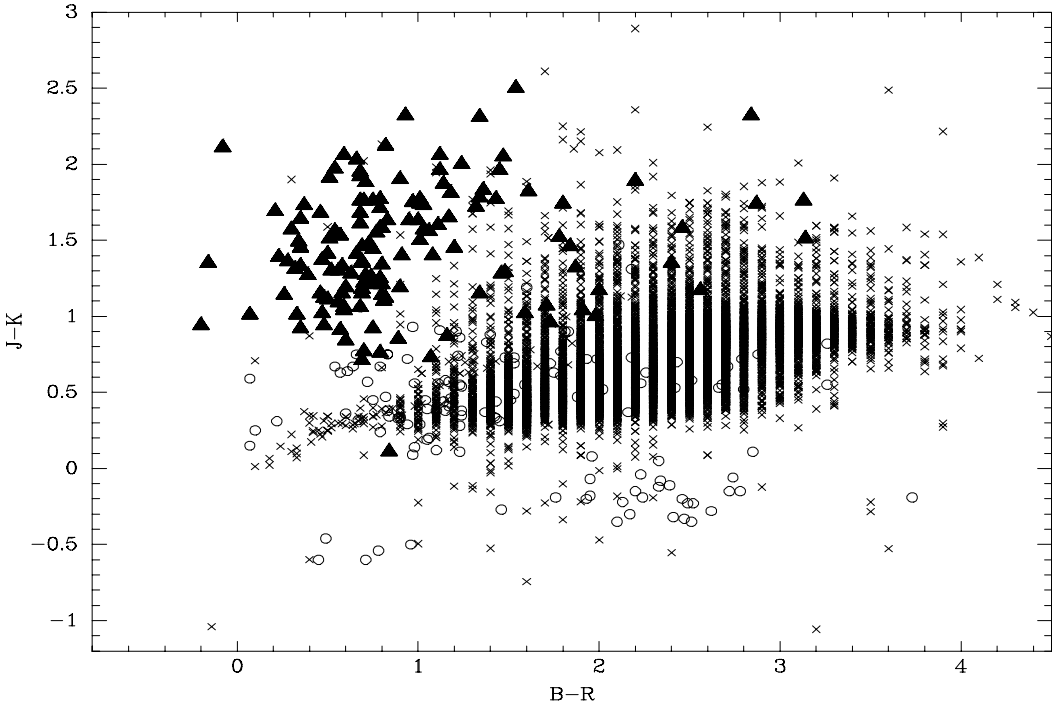,height=10cm}
\caption{Optical and near-IR colours of the Parkes sources (triangles) 
compared to photometry of 6400 high galactic latitude point sources 
drawn from the 2MASS survey (crosses) and sources with $K<22$ from the EIS
Hubble Deep Field data release (circles).
\label{all_select}}
\end{center}
\end{figure}

In Fig~\ref{all_select}, we compare the optical and near-IR colours
of the Parkes sources against the colours of high galactic latitude
point sources drawn from the Two-Micron All Sky Survey (2MASS, $K<15$) and
ESO Imaging Survey (EIS, $K<22$). The `Main Sequence' sources, both red and
blue, are clearly separated from the foreground objects. This separation
is due to their power-law spectral energy distributions: as compared to the
convex spectral energy distributions of stars and galaxies, the quasars
have excess flux in $B$ and/or $K$. This selection technique is
similar to the `KX' technique proposed by Warren, Hewett \& Foltz (1999).
Unfortunately, the very red sources lying below the `main sequence' have
colours within the stellar locus and will be hard to find.

\begin{figure}
\begin{center}
\psfig{file=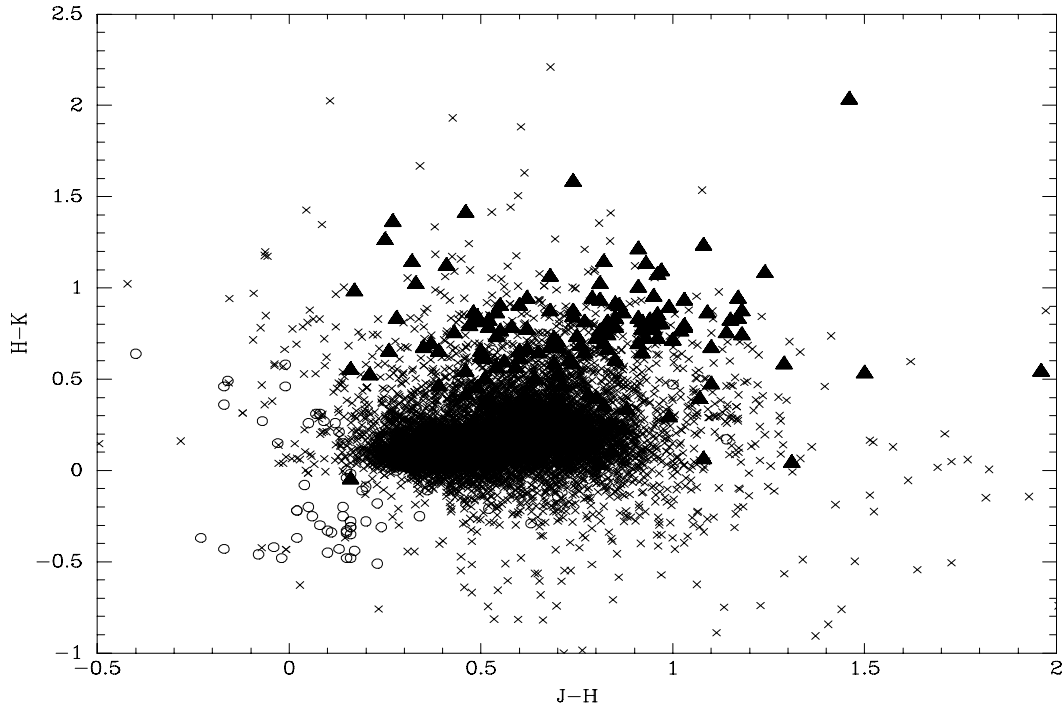,height=10cm}
\caption{Near-IR colours of the Parkes sources (triangles) compared to
photometry of 6400 high galactic latitude point sources drawn from
the 2MASS survey (crosses) and sources with $K<22$ from the EIS
Hubble Deep Field data release (circles).
\label{ir_select}}
\end{center}
\end{figure}

\begin{figure}
\begin{center}
\psfig{file=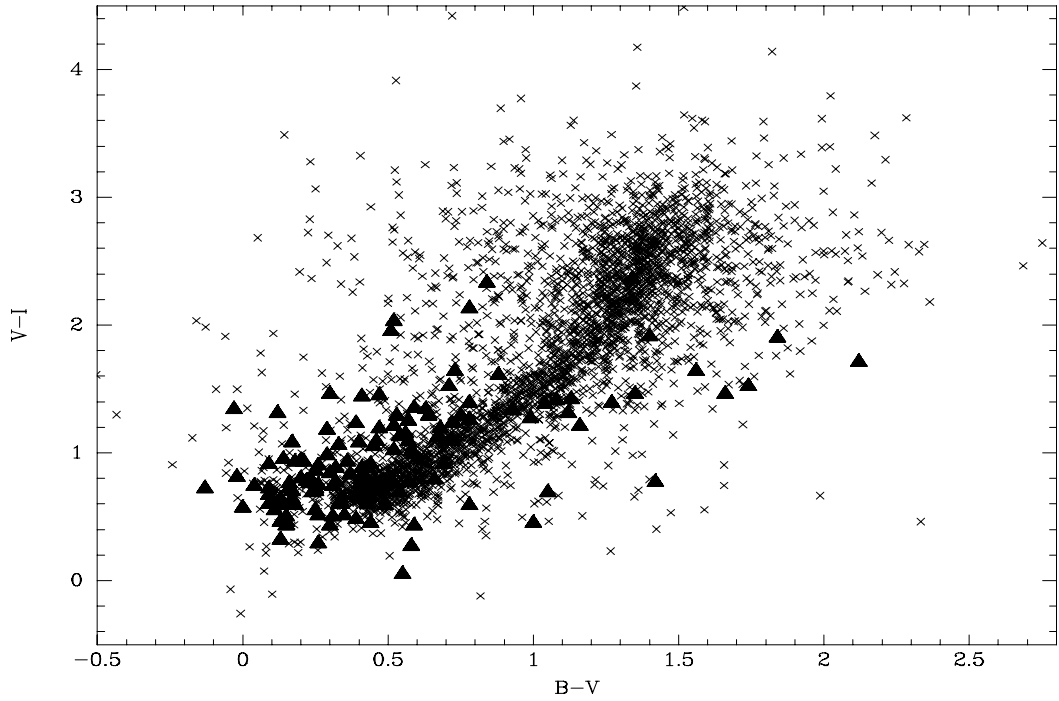,height=10cm}
\caption{Optical colours of the Parkes sources (triangles) compared to
photometry of 3200 high galactic latitude point sources drawn from
the EIS wide survey (crosses).
\label{opt_select}}
\end{center}
\end{figure}

Can red quasars be identified purely on the basis of their near-IR
colours? In Fig~\ref{ir_select}, we show that most of the Parkes
quasars lie in regions of the near-IR colour-colour plot with
substantial stellar contamination, but that the reddest move away
from the stellar locus, and could be detectable in the IR alone.
Fig~\ref{opt_select} shows that purely optical colour selection is not
likely to be effective.

\section{Conclusions}

The Parkes quasars can, we conclude, be crudely divided into
three populations:

\begin{enumerate}

\item {\bf The `Main Sequence'}:
$\sim 90$\% of the Parkes sources have approximately power-law
spectral energy distributions, with spectral indices $\alpha$
($F_{\nu} \propto \nu^{\alpha}$) in the range $0 > \alpha > -2$.
The nature of these sources is discussed by Whiting et al.

\item {\bf Very Red Sources}: These sources, which comprise
$\sim 10$\% of the Parkes sample, are characterised by much redder
continuum slopes in the optical than in the IR. They tend to
have relatively steep radio spectra. Half these sources are 
radio galaxies, while most of the remainder are highly dust-reddened
quasars. The undetected sources are probably high redshift members of
this class.

\item {\bf Oddballs}: Roughly 2\% of the Parkes sample defy this
categorisation.

\end{enumerate}

The `main sequence' sources, both red and blue, should be easily
detectable in combined near-IR and optical QSO surveys, due to their
excess flux in the $K$ and/or $B$ bands.

\section*{Acknowledgements}

We wish to thank Mike Bessell and Peter MacGregor for their help
with the details of the photometry, and Tori Ibbetson for her
assistance with the observations. This publication makes use of data
products from the Two Micron All Sky Survey, which is a joint project
of the University of Massachusetts and the Infrared Processing and
Analysis Center, funded by the National Aeronautics and Space
Administration and the National Science Foundation, and of catalogues
from the ESO Imaging Survey, obtained from observations with the ESO New
Technology Telescope at the La Silla observatory under program-ID 
Nos 59.A-9005(A) and 60.A-9005(A).

\section*{References}

\reference Bersanelli, M., Bouchet, P., \& Falomo, R. 1991,
A \& A, 252, 854

\reference Bessell, M. S., Castelli, F., \& Plez, B. 1998,
A \& A, 333, 231

\reference Carter, B. S., \& 
Meadows, V. S. 1995, MNRAS, 276, 734

\reference Drinkwater, M. J., Webster, R. L., Francis, P. J., Condon, J. J.,
Ellison, S. L., Jauncey, D. L., Lovell, J., Peterson, B. A.,
\& Savage, A. 1997, MNRAS, 284, 85 

\reference Francis, P. J., Hewett, P. C., Foltz, C. B., \& Chaffee, C. B.,
1991, ApJ 373, 465

\reference Francis, P. J. 1996, 
Publ. Astron. Soc. Australia, 13, 212

\reference Graham J. A. 1982, PASP 94,
265

\reference Heidt, J. \& Wagner, S.J. 1996, A \& A, 305, 42

\reference Ledden, S.E. \& O'Dell, S.L. 1983, ApJ, 270, 434

\reference MacKenty, J. W. et al, 1997, ``NICMOS Instrument Handbook'',
Version 2.0, (Baltimore: STScI)

\reference Masci, F. J., Webster, R. L. \& Francis, P. J. 1998,
MNRAS 301, 975

\reference McCarthy, P. 1992, Ann Rev A \& A, 31, 639

\reference McGregor, P., Hart, J., 
Downing, M., Hoadley, D., \& Bloxham, G. 1994, in Infrared Astronomy with 
Arrays: The Next Generation, ed. I. S. McLean (Kluwer: Dordrecht), p. 299

\reference Morris S. L., Weymann R.
J., Anderson S. F., Hewett P. C., Foltz C. B., Chaffee F. H. \&
Francis P. J. 1991, AJ, 102, 1627

\reference Neugebauer, G., Green, R. F., Matthews, K., Schmidt, M.,
Soifer, B. T., \& Bennet, J. 1987, ApJS, 63, 615

\reference Rieke, G.H., Lebofsky, M.J. \& Wisniewski, W.A. 1982,
ApJ 263, 73

\reference Schlegel, D. J., Finkbeiner, D. P., \& Davis, M. 1998,
ApJ 500, 525

\reference Serjeant, S., \& Rawlings, S. 1997, Nature, 379, 304

\reference Stickel, M., Rieke, G.H., K\"{u}hr, H. \& Rieke, M.J. 1996,
ApJ, 468, 556

\reference Wagner, S.J., Sanchez-Pons, F., Quirrenbach, A., \& Witzel, A.
1990, A \& A, 235 L1

\reference Warren, S.J., Hewett, P.C. \& Foltz, C.B. 1999, MNRAS in press
(astro-ph/9911064).

\reference Webster, R. L., Francis, P. J., Peterson, B. A.,
Drinkwater, M. J., \& Masci, F. J. 1995, Nature, 375, 469

\reference Whiting, M.T., Webster, R.L. \& Francis, P.J. 2000,
MNRAS submitted.

\reference Wilkes, B. J., Wright, A. E., Jauncey, D. L. \&
Peterson, B. A., 1983, PASA, 5, 2

\reference Wills, B., Netzer, H., \& Wills, D. 1985, ApJ 288, 94

\end{document}

\setcounter{table}{1}

\begin{deluxetable}{lcccrcrcrcrcrcrcr}
\scriptsize
\tablewidth{0pt}
\tablecolumns{18}
\tablecaption{Observation Dates and Exposure Times}
\tablehead{
\colhead{ ~ } &
\colhead{Redshift} &
\colhead{Morph-} &
\multicolumn{2}{c}{B} &
\multicolumn{2}{c}{V} &
\multicolumn{2}{c}{R} &
\multicolumn{2}{c}{I} &
\multicolumn{2}{c}{J} &
\multicolumn{2}{c}{H} &
\multicolumn{2}{c}{K$_n$} \\
\colhead{Name} &
\colhead{z} &
\colhead{ology} &
\multicolumn{14}{c}{Letters indicate observation dates, numbers are exposure 
times (seconds)}
}
\startdata
PKS 0036$-$216 & \nodata & galaxy & W &  600 & W &  300 & W &  300 & W &  300 & X &  240 & X &  240 & X & 1200 \nl
PKS 0038$-$020 & 1.178  &  point & U &  360 & U &  360 & U &  360 & U &  360 & X &  240 & X &  240 & X & 1200 \nl
PKS 0048$-$071 & 1.974  &  faint & W &  300 & W &  300 & W &  300 & W &  300 & X & 1200 & X & 1200 & X & 1200 \nl
PKS 0048$-$097 & \nodata &  point & W &  180 & W &  180 & W &  180 & W &  180 & X &  240 & X &  240 & X & 1200 \nl
PKS 0048$-$427 & 1.749  &  point & W &  300 & W &  300 & W &  300 & W &  300 & X &  240 & X &  240 & X & 1200 \nl
PKS 0056$-$001 & 0.717  &  point & U &  300 & U &  300 & U &  300 & \nodata & \nodata & X &  240 & X &  240 & X & 1200 \nl
PKS 0131$-$001 & 0.879  &  faint & M & 1200 & M &  600 & M &  600 & M &  600 & N & 1200 & N & 1200 & N & 1200 \nl
PKS 0153$-$410 & 0.226  & galaxy & V &  300 & V &  180 & V &  180 & V &  180 & X &  240 & X &  240 & X & 1200 \nl
PKS 0202$-$172 & 1.740  &  point & V &  180 & V &  180 & V &  180 & V &  180 & X &  240 & X &  240 & X & 1200 \nl
PKS 0213$-$026 & 1.178  &  faint & V &  600 & V &  600 & V &  600 & V &  600 & X &  240 & X &  240 & X & 1200 \nl
PKS 0216$+$011 & 1.610  &  faint & V &  300 & V &  300 & V &  300 & V &  300 & X &  480 & X & 1200 & X & 1200 \nl
PKS 0220$-$349 & 1.490  &  faint & V &  300 & V &  300 & V &  300 & V &  300 & X & 1200 & X & 1200 & X & 1200 \nl
PKS 0221$+$067 & 0.510  &  point & V &  300 & V &  300 & V &  300 & V &  300 & Z &  240 & Z &  240 & Z &  240 \nl
PKS 0226$-$038 & 2.066  &  point & V &  180 & V &  180 & V &  180 & V &  180 & Z &  240 & Z &  240 & Z & 1200 \nl
PKS 0229$-$398 & 1.646  &  point & V &  600 & V &  600 & V &  600 & V &  600 & Z & 1200 & X & 1200 & X & 1200 \nl
PKS 0232$-$042 & 1.437  &  point & V &  180 & V &  180 & V &  180 & V &  180 & Z &  240 & Z &  240 & Z & 1200 \nl
PKS 0237$+$040 & 0.978  &  point & W &  300 & W &  180 & W &  180 & W &  180 & Z &  240 & Z &  240 & Z & 1200 \nl
PKS 0238$-$084 & 0.005  & galaxy\tablenotemark{a} & W &   60 & W &   60 & W &   60 & W &   60 & Z &  300 & Z &  300 & Z &  300 \nl
PKS 0240$-$060 & 1.800  &  point & W &  180 & W &  180 & W &  180 & W &  180 & Z &  240 & Z &  240 & Z &  240 \nl
PKS 0240$-$217 & 0.314  & galaxy\tablenotemark{b} & W &  180 & W &  180 & W &  180 & W &  180 & \nodata & \nodata & Z &  240 & Z &  300 \nl
PKS 0256$+$075 & 0.895  &  point & W &  960 & W &  600 & W &  600 & W &  600 & \nodata & \nodata & \nodata & \nodata & \nodata &    0 \nl
PKS 0301$-$243 & \nodata &  point & W &  180 & W &  180 & W &  180 & W &  180 & \nodata & \nodata & \nodata & \nodata & \nodata &    0 \nl
PKS 0316$-$444 & 0.076  & galaxy & W &  180 & W &  180 & W &  180 & W &  180 & \nodata & \nodata & \nodata & \nodata & \nodata &    0 \nl
PKS 0537$-$441 & 0.893  &  point & A &  300 & A &  300 & A &  300 & A &  300 & F &  240 & E &  240 & F &  240 \nl
PKS 0829$+$046 & \nodata &  point & A &  300 & A &  300 & A &  300 & A &  300 & F &  240 & E &  240 & F &  240 \nl
PKS 0912$+$029 & 0.427  &  point & C &  600 & C &  300 & C &  300 & C &  300 & E & 1200 & E & 1200 & E & 2240 \nl
PKS 0922$+$005 & 1.717  &  point & \nodata & \nodata & \nodata & \nodata & D &  300 & D &  300 & \nodata & \nodata 
& \nodata & \nodata & \nodata &    0 \nl
PKS 1016$-$311 & 0.794  &  point & B &  300 & B &  300 & B &  300 & B &  300 & F &  240 & F &  240 & F &  240 \nl
PKS 1020$-$103 & 0.197  &  point & A &  300 & A &  300 & A &  300 & A &  300 & F &  240 & F &  240 & F &  240 \nl
PKS 1021$-$006 & 2.549  &  point & B &  300 & B &  300 & B &  300 & B &  300 & F &  240 & \nodata & \nodata & \nodata &    0 \nl
PKS 1032$-$199 & 2.198  &  point & B &  300 & B &  300 & B &  300 & B &  300 & F &  240 & F &  240 & F & 1200 \nl
PKS 1034$-$293 & 0.312  &  point & B &  300 & B &  300 & B &  300 & B &  300 & F &  240 & F &  240 & F &  240 \nl
PKS 1036$-$154 & 0.525  &  point & C &  600 & C &  300 & C &  300 & C &  300 & F &  240 & E &  240 & F &  240 \nl
PKS 1038$+$064 & 1.264  &  point & A &  300 & A &  300 & A &  300 & A &  300 & \nodata & \nodata & \nodata & \nodata & \nodata &    0 \nl
PKS 1042$+$071 & 0.698  &  point & B &  600 & B &  300 & B &  300 & B &  300 & \nodata & \nodata & \nodata & \nodata & \nodata &    0 \nl
PKS 1045$-$188 & 0.595  &  point & B &  600 & B &  300 & B &  300 & B &  300 & F &  240 & F &  240 & F & 2240 \nl
PKS 1048$-$313 & 1.429  &  point & B &  600 & B &  300 & B &  300 & B &  300 & \nodata & \nodata & \nodata & \nodata & \nodata &    0 \nl
PKS 1055$-$243 & 1.086  &  point & B &  300 & B &  300 & B &  300 & B &  300 & \nodata & \nodata & \nodata & \nodata & \nodata &    0 \nl
PKS 1055$+$018 & 0.888  &  point & B &  300 & B &  300 & B &  300 & B &  300 & \nodata & \nodata & \nodata & \nodata & \nodata &    0 \nl
PKS 1101$-$325 & 0.355  &  point & A &  300 & A &  300 & A &  300 & A &  300 & F &  240 & E &  240 & F &  240 \nl
PKS 1107$-$187 & 0.497  &  faint & D & 2400 & D & 1800 & D & 1200 & D & 1200 & E & 1200 & E & 1200 & E & 1200 \nl
PKS 1110$-$217 & \nodata &  faint & C & 2400 & C & 2400 & C & 3600 & C & 1800 & E & 1200 & E & 1200 & E & 1200 \nl
PKS 1133$-$172 & 1.024  &  faint & D & 1800 & D & 1800 & D & 1200 & D & 1200 & E &  240 & E &  240 & E &  240 \nl
PKS 1136$-$135 & 0.557  &  point & A &  300 & A &  300 & A &  300 & A &  300 & F &  240 & F &  240 & F & 1200 \nl
PKS 1156$-$094 & \nodata &  faint & C &  600 & C &  600 & C &  600 & C &  600 & F &  240 & F &  240 & F & 1200 \nl
PKS 1244$-$255 & 0.638  &  point & A &  300 & A &  300 & A &  300 & A &  300 & F &  240 & F &  240 & F &  240 \nl
PKS 1256$-$229 & 1.365  &  point & A &  300 & A &  300 & A &  300 & A &  300 & F &  240 & F &  240 & F &  240 \nl
PKS 1313$-$333 & 1.210  &  point & A &  300 & A &  300 & A &  300 & A &  300 & F &  240 & E &  240 & F &  240 \nl
PKS 1330$+$022 & 0.216  & galaxy & A &  600 & A &  300 & A &  300 & A &  300 & F &  240 & F &  240 & F &  240 \nl
PKS 1353$-$341 & 0.223  & galaxy\tablenotemark{c} & A & 1200 & A &  300 & A &  300 & A &  300 & F &  240 & F &  240 & F &  240 \nl
PKS 1404$-$342 & 1.122  &  point & K &  300 & K &  300 & K &  300 & K &  300 & O &  240 & O &  240 & O &  240 \nl
PKS 1411$+$094 & 0.162  & galaxy\tablenotemark{d} & K &  300 & K &  300 & K &  300 & K &  300 & O &  240 & O &  240 & O & 1200 \nl
PKS 1430$-$155 & 1.573  &  faint & L & 1800 & L & 1800 & L & 1200 & L & 1200 & O & 1200 & O & 2400 & O & 1200 \nl
PKS 1430$-$178 & 2.326  &  point & J &  300 & J &  300 & J &  300 & J &  300 & P &  240 & P &  240 & P &  240 \nl
PKS 1435$-$218 & 1.187  &  point & J &  300 & J &  300 & J &  300 & J &  300 & P &  240 & P &  240 & P &  240 \nl
PKS 1437$-$153 & 0.254  &  point & J &  300 & J &  300 & J &  300 & J &  300 & P &  240 & P &  240 & P &  240 \nl
PKS 1438$-$347 & 1.159  &  point & J &  300 & J &  300 & J &  300 & J &  300 & P &  240 & P &  240 & P &  240 \nl
PKS 1450$-$338 & 0.368  &  faint & D & 2400 & D & 1200 & D &  600 & D &  600 & E &  240 & E &  240 & E &  240 \nl
PKS 1454$-$060 & 1.249  &  point & I &  300 & I &  300 & I &  300 & I &  300 & P &  240 & P &  240 & P &  240 \nl
PKS 1456$+$044 & 0.394  &  point & \nodata & \nodata & \nodata & \nodata & C &  900 & C &  900 & E & 1200 & E & 1200 & F & 1200 \nl
PKS 1504$-$166 & 0.876  &  point & I &  600 & I &  300 & I &  300 & I &  300 & P &  240 & P &  240 & P &  240 \nl
PKS 1508$-$055 & 1.185  &  point & I &  300 & I &  300 & I &  300 & I &  300 & P &  240 & P &  240 & P &  240 \nl
PKS 1509$+$022 & 0.219  & galaxy\tablenotemark{e} & I &  300 & I &  300 & \nodata & \nodata & \nodata & \nodata & P &  240 & P &  240 & P &  240 \nl
PKS 1510$-$089 & 0.362  &  point & A &  300 & A &  300 & A &  300 & A &  300 & E &  240 & E &  240 & E &  240 \nl
PKS 1511$-$100 & 1.513  &  point & B &  300 & B &  300 & B &  300 & B &  300 & F &  240 & F &  240 & F &  240 \nl
PKS 1511$-$210 & 1.179  &  point & L &  900 & L &  600 & L &  600 & L &  600 & Q &  240 & Q &  240 & Q &  240 \nl
PKS 1514$-$241 & 0.049  & galaxy\tablenotemark{f} & B &  300 & B &   60 & B &   60 & B &   60 & P &  240 & P &  240 & P &  240 \nl
PKS 1518$+$045 & 0.052  & galaxy\tablenotemark{g} & K &  300 & K &  300 & K &  300 & K &  300 & P &  240 & P &  240 & P &  240 \nl
PKS 1519$-$273 & 0.510  &  point & K &  300 & K &  300 & K &  300 & K &  300 & P &  240 & P &  240 & P &  240 \nl
PKS 1532$+$016 & 1.435  &  point & K &  300 & K &  300 & K &  300 & K &  300 & P &  240 & P &  240 & P &  240 \nl
PKS 1535$+$004 & 3.497  &  faint & M &  600 & \nodata & \nodata & M & 2400 & M & 2400 & \nodata & \nodata & R & 3600 
& \nodata &    0 \nl
PKS 1542$+$042 & 2.184  &  point & J &  300 & J &  300 & J &  300 & J &  300 & Q &  240 & Q &  240 & Q &  240 \nl
PKS 1546$+$027 & 0.415  &  point & A &  300 & A &  300 & A &  300 & A &  300 & F &  240 & E &  240 & F &  240 \nl
PKS 1548$+$056 & 1.422  &  point & K &  900 & K &  300 & K &  300 & K &  300 & Q &  240 & Q &  240 & Q &  240 \nl
PKS 1550$-$269 & 2.145  &  point & B &  300 & B &  300 & B &  300 & B &  300 & F &  240 & F &  240 & F & 1200 \nl
PKS 1555$-$140 & 0.097  & galaxy\tablenotemark{h} & J &  300 & J &  300 & J &  300 & J &  300 & Q &  240 & Q &  240 & Q &  240 \nl
PKS 1555$+$001 & 1.770  &  faint & M &  600 & M &  600 & M &  900 & M &  300 & Q &  240 & Q &  240 & Q &  240 \nl
PKS 1556$-$245 & 2.818  &  point & I &  360 & I &  360 & I &  300 & I &  300 & Q &  240 & Q &  240 & Q &  240 \nl
PKS 1601$-$222 & \nodata &  faint & \nodata & \nodata & \nodata & \nodata & L & 2400 & L & 2400 & \nodata & \nodata & 
Q & 3600 & Q & 3600 \nl
PKS 1602$-$001 & 1.624  &  point & J &  300 & J &  300 & J &  300 & J &  300 & O &  240 & O &  240 & P & 1200 \nl
PKS 1614$+$051 & 3.217  &  point & K &  900 & K &  600 & K &  600 & K &  900 & O &  240 & O &  240 & O & 1200 \nl
PKS 1615$+$029 & 1.341  &  point & A &  300 & AD & 2100 & AD &  900 & AD &  900 & E & 1200 & E & 1200 & E & 1200 \nl
PKS 1616$+$063 & 2.088  &  point & J &  300 & J &  300 & J &  300 & J &  300 & O &  240 & O &  240 & O & 1200 \nl
PKS 1635$-$035 & 2.856  &  point & D &  600 & D &  600 & D &  600 & D &  600 & F & 2240 & F & 2240 & F & 1200 \nl
PKS 1648$+$015 & \nodata &  faint & C &  600 & C &  600 & C &  600 & C &  600 & E & 1200 & E & 1200 & E & 1200 \nl
PKS 1649$-$062 & \nodata &  faint & \nodata & \nodata & \nodata & \nodata & C & 3000 & C & 2400 & F & 2240 & F & 2240 & F & 2240 \nl
PKS 1654$-$020 & 2.000  &  faint & M & 1800 & M & 1800 & M & 1800 & M & 1800 & Q & 1200 & Q & 1200 & Q & 1200 \nl
PKS 1655$+$077 & 0.621  &  point & V &  600 & V &  600 & V &  300 & V &  300 & Y &  240 & Y &  240 & Y &  240 \nl
PKS 1656$+$053 & 0.887  &  point & I &  300 & I &  300 & I &  300 & I &  300 & P &  240 & P &  240 & P &  240 \nl
PKS 1705$+$018 & 2.576  &  point & I &  300 & I &  300 & I &  300 & I &  300 & Q &  240 & Q &  240 & Q & 1200 \nl
PKS 1706$+$006 & 0.449  &  faint & V & 1200 & V &  600 & V &  300 & V &  300 & X &  240 & X &  240 & X &  240 \nl
PKS 1725$+$044 & 0.296  &  point & A &  600 & A &  300 & A &  300 & A &  300 & E & 1200 & E & 1200 & E & 1200 \nl
PKS 1732$+$094 & \nodata &  faint & M & 1800 & L & 1200 & L & 1200 & L & 1200 & P & 1200 & P & 1200 & P & 1200 \nl
PKS 1933$-$400 & 0.965  &  point & A &  300 & A &  300 & A &  300 & A &  300 & F &  240 & F &  240 & F &  240 \nl
PKS 1953$-$325 & 1.242  &  point & I &  300 & I &  300 & I &  300 & I &  300 & N &  240 & N &  240 & N &  240 \nl
PKS 1954$-$388 & 0.626  &  point & I &  300 & I &  300 & I &  300 & I &  300 & N &  240 & N &  240 & N &  240 \nl
PKS 1958$-$179 & 0.652  &  point & I &  300 & I &  300 & I &  300 & I &  300 & N &  240 & N &  240 & N &  240 \nl
PKS 2000$-$330 & 3.783  &  point & I &  300 & I &  300 & I &  300 & I &  300 & P & 1200 & P & 1200 & P & 1200 \nl
PKS 2002$-$185 & 0.859  &  point & B &  600 & B &  300 & B &  300 & B &  300 & F &  240 & E &  240 & E &  240 \nl
PKS 2004$-$447 & 0.240  &  point & A &  300 & A &  300 & A &  300 & A &  300 & F &  240 & F &  240 & E &  240 \nl
PKS 2008$-$159 & 1.178  &  point & I &  300 & I &  300 & I &  300 & I &  300 & N &  240 & N &  240 & N &  240 \nl
PKS 2021$-$330 & 1.471  &  point & J &  300 & I &  300 & I &  300 & I &  300 & NQ &  480 & NQ &  480 & NQ &  720 \nl
PKS 2022$-$077 & 1.388  &  point & J &  600 & J &  600 & J &  600 & J &  600 & N &  240 & N &  240 & N &  240 \nl
PKS 2037$-$253 & 1.574  &  point & J &  300 & J &  300 & J &  300 & J &  300 & O &  240 & O &  240 & O &  240 \nl
PKS 2044$-$168 & 1.937  &  point & J &  300 & J &  300 & J &  300 & J &  300 & N &  240 & N &  240 & N &  240 \nl
PKS 2047$+$098 & \nodata &  faint & \nodata & \nodata & \nodata & \nodata & VW & 1800 & W & 4200 & \nodata & \nodata & X & 3600 
& \nodata &    0 \nl
PKS 2053$-$044 & 1.177  &  point & J &  300 & J &  300 & J &  300 & J &  300 & O &  240 & O &  240 & O &  240 \nl
PKS 2056$-$369 & \nodata &  faint & \nodata & \nodata & M & 2400 & L & 1800 & L & 1800 & O & 1200 & O & 1200 & OQ & 2400 \nl
PKS 2058$-$135 & 0.029  & galaxy\tablenotemark{i} & J &  120 & J &   30 & J &   30 & J &   30 & Q &  240 & Q &  240 & Q &  240 \nl
PKS 2058$-$297 & 1.492  &  point & J &  300 & J &  300 & J &  300 & J &  300 & P &  480 & P &  480 & P &  240 \nl
PKS 2059$+$034 & 1.012  &  point & J &  300 & J &  300 & J &  300 & J &  300 & Q &  240 & Q &  240 & Q &  240 \nl
PKS 2106$-$413 & 1.055  &  point & J &  300 & J &  300 & J &  300 & J &  300 & Q &  240 & Q &  240 & Q &  240 \nl
PKS 2120$+$099 & 0.932  &  point & T &  300 & T &  300 & T &  300 & T &  300 & X &  240 & X &  240 & X &  240 \nl
PKS 2121$+$053 & 1.941  &  point & J &  300 & J &  300 & J &  300 & J &  300 & Q &  240 & Q &  240 & Q &  240 \nl
PKS 2126$-$158 & 3.266  &  point & L &  600 & L & 1860 & L & 1860 & L & 1800 & P & 1200 & P & 1200 & P & 2240 \nl
PKS 2127$-$096 & 0.780  &  faint & T &  600 & T &  300 & T &  300 & T &  300 & X &  240 & X &  240 & X &  240 \nl
PKS 2128$-$123 & 0.499  &  point & K &  300 & K &  300 & K &  300 & K &  300 & O & 1200 & O & 1200 & O & 1240 \nl
PKS 2131$-$021 & 1.285  &  point & K &  900 & K &  300 & K &  300 & K &  300 & Q &  240 & Q &  240 & Q &  240 \nl
PKS 2134$+$004 & 1.937  &  point & K &  300 & K &  300 & K &  300 & K &  300 & Q &  240 & Q &  240 & Q &  240 \nl
PKS 2135$-$248 & 0.821  &  point & K &  300 & K &  300 & K &  300 & K &  300 & Q &  240 & Q &  240 & Q &  240 \nl
PKS 2140$-$048 & 0.344  &  point & K &  300 & K &  300 & K &  300 & K &  300 & Q &  240 & Q &  240 & Q &  240 \nl
PKS 2143$-$156 & 0.698  &  point & K &  300 & K &  300 & K &  300 & K &  300 & Q &  240 & Q &  240 & Q &  240 \nl
PKS 2144$+$092 & 1.113  &  point & T &  300 & T &  300 & T &  300 & T &  300 & X &  240 & X &  240 & X &  240 \nl
PKS 2145$-$176 & 2.130  &  point & T &  300 & T &  300 & T &  300 & T &  300 & X &  240 & X &  240 & X &  240 \nl
PKS 2145$+$067 & 0.999  &  point & V &  120 & V &  120 & V &  120 & U &  300 & X &  240 & X &  240 & X &  240 \nl
PKS 2149$-$307 & 2.345  &  point & T &  300 & T &  300 & T &  300 & T &  300 & X &  240 & X &  240 & X &  240 \nl
PKS 2149$+$056 & 0.740  &  faint & M & 1800 & M & 1800 & M & 1200 & M &  600 & Q &  240 & Q &  240 & Q &  240 \nl
PKS 2149$+$069 & 1.364  &  point & V &  120 & V &  120 & V &  120 & V &  120 & X &  240 & X &  240 & X &  240 \nl
PKS 2155$-$152 & 0.672  &  point & M & 1200 & M & 1200 & M & 1800 & M & 1800 & Q & 1200 & Q & 1200 & Q &  960 \nl
PKS 2200$-$238 & 2.120  &  point & T &  300 & T &  300 & T &  300 & T &  300 & Y &  240 & Y &  240 & Y & 1200 \nl
PKS 2203$-$188 & 0.619  &  point & T &  300 & T &  300 & T &  300 & T &  300 & Y &  240 & Y &  240 & Y & 1200 \nl
PKS 2206$-$237 & 0.086  & galaxy\tablenotemark{j} & V &  120 & V &  120 & V &  120 & V &  120 & Y &  240 & Y &  240 & Y &  240 \nl
PKS 2208$-$137 & 0.392  &  point & V &  120 & V &  120 & V &  120 & V &  120 & Y &  240 & Y &  240 & Y & 1200 \nl
PKS 2210$-$257 & 1.833  &  point & V &  120 & V &  120 & V &  120 & V &  120 & Y &  240 & Y &  240 & Y & 1320 \nl
PKS 2212$-$299 & 2.703  &  point & V &  120 & V &  120 & V &  120 & V &  120 & Y &  240 & Y &  240 & Y & 1200 \nl
PKS 2215$+$020 & 3.572  &  point & V &  600 & V &  600 & V &  300 & V &  300 & Y &  240 & Y &  360 & Y &  240 \nl
PKS 2216$-$038 & 0.901  &  point & L &  600 & L &  600 & L & 1860 & L & 1800 & P &  600 & P & 1200 & P & 1200 \nl
PKS 2223$-$052 & 1.404  &  point & V &  120 & V &  120 & V &  120 & V &  120 & Z &  240 & Z &  240 & Z & 1200 \nl
PKS 2227$-$088 & 1.561  &  point & V &  120 & V &  120 & V &  120 & V &  120 & Z &  240 & Z &  240 & Z &  240 \nl
PKS 2227$-$399 & 0.323  &  point & V &  120 & V &  120 & V &  120 & V &  120 & Z &  240 & Z &  240 & Z &  240 \nl
PKS 2229$-$172 & 1.780  &  faint & V &  600 & V &  600 & V &  300 & V &  300 & Z &  480 & Z &  240 & Z & 1200 \nl
PKS 2233$-$148 & 0.609  &  point & V &  600 & V &  300 & V &  300 & V &  300 & Z &  240 & Z &  240 & Z & 1200 \nl
PKS 2239$+$096 & 1.707  &  point & W &  300 & W &  300 & W &  300 & W &  300 & Z &  240 & Z &  240 & Z &  240 \nl
PKS 2240$-$260 & 0.774  &  point & W &  180 & W &  180 & W &  180 & W &  180 & Z &  240 & Z &  240 & Z & 1260 \nl
PKS 2243$-$123 & 0.630  &  point & W &  180 & W &  180 & W &  180 & W &  180 & Z &  240 & Z &  240 & Z &  240 \nl
PKS 2245$-$328 & 2.268  &  point & W &  180 & W &  180 & W &  180 & W &  180 & Z &  240 & Z &  240 & Z &  240 \nl
PKS 2245$+$029 & \nodata  &  point & W & 1200 & W &  600 & W &  600 & W &  600 & X &  300 & X &  240 & X &  240 \nl
PKS 2252$-$090 & \nodata  &  faint & W & 1800 & W & 1800 & W &  600 & W &  600 & X &  240 & X &  240 & X &  240 \nl
PKS 2312$-$319 & 1.323  &  point & J &  300 & J &  300 & J &  300 & J &  300 & O &  240 & O &  240 & O & 1200 \nl
PKS 2313$-$438 & 1.847  &  point & J &  300 & J &  300 & J &  300 & J &  300 & Q &  240 & Q &  240 & Q & 1200 \nl
PKS 2314$-$409 & 2.448  &  point & J &  300 & J &  300 & J &  300 & J &  300 & Q &  240 & Q &  240 & Q & 1200 \nl
PKS 2329$-$415 & 0.671  &  point & J &  300 & J &  300 & J &  300 & J &  300 & Q &  240 & Q &  240 & Q & 1200 \nl
PKS 2337$-$334 & \nodata  &  faint & L &  600 & L &  600 & L &  600 & L & 1200 & O &  960 & O &  600 & O &  660 \nl
PKS 2344$-$192 & \nodata  &  faint &L M & 1800 & \nodata & \nodata & L &  600 & L &  600 & PQ & 2400 & PQ & 2400 & PQ & 2400 \nl
PKS 2345$-$167 & 0.576  &  point & K &  300 & K &  300 & K &  300 & K &  300 & Q &  240 & Q &  240 & Q & 1200 \nl
PKS 2351$-$154 & 2.675  &  point & K &  300 & K &  300 & K &  300 & K &  300 & Q &  240 & Q &  240 & Q &  240 \nl
PKS 2354$-$117 & 0.960  &  point & K &  300 & K &  300 & K &  300 & K &  300 & Q &  240 & Q &  240 & Q & 1200 \nl
\tablevspace{2mm}
L 2110$-$4509 & 0.555   &  point & U &  300 & U &  300 & U &  300 & U &  300 & Y &  240 & Y &  240 & Y &  240 \nl
L 2111$-$4335 & 1.708   &  point & U &  300 & U &  300 & U &  300 & U &  300 & Y &  240 & Y &  240 & Y &  240 \nl
L 2111$-$4506 & 1.376   &  point & U &  300 & U &  300 & U &  300 & U &  300 & Y &  240 & Y &  240 & Y &  240 \nl
L 2113$-$4305 & 1.249   &  point & U &  300 & U &  300 & U &  300 & U &  300 & Y &  240 & Y &  240 & Y &  240 \nl
L 2113$-$4345 & 2.053   &  point & U &  300 & U &  300 & U &  300 & U &  300 & Y &  240 & Y &  240 & Y &  240 \nl
L 2113$-$4538 & 0.946   &  point & U &  300 & U &  300 & U &  300 & U &  300 & \nodata & \nodata & \nodata & \nodata & \nodata &    0 \nl
L 2114$-$4335 & 1.318   &  point & W &  180 & W &  180 & W &  180 & W &  180 & \nodata & \nodata & \nodata & \nodata & \nodata &    0 \nl
L 2114$-$4346 & 2.041   &  point & W &  180 & W &  180 & W &  180 & W &  180 & \nodata & \nodata & \nodata & \nodata & \nodata &    0 \nl
L 2114$-$4501 & 0.597   &  point & W &  180 & W &  180 & W &  180 & W &  180 & \nodata & \nodata & \nodata & \nodata & \nodata &    0 \nl
L 2116$-$4439 & 1.480   &  point & W &  180 & W &  180 & W &  180 & W &  180 & \nodata & \nodata & \nodata & \nodata & \nodata &    0 \nl
L 2118$-$4702 & 1.332   &  point & W &  180 & W &  180 & W &  180 & W &  180 & \nodata & \nodata & \nodata & \nodata & Z &  240 \nl
L 2119$-$4415 & 0.728   &  point & W &  180 & W &  180 & W &  180 & W &  180 & Z &  240 & Z &  240 & Z &  240 \nl
\enddata
\tablenotetext{a}{Use circular photometric aperture of radius 45$\arcsec$}
\tablenotetext{b}{Aperture radius 10$\arcsec$}
\tablenotetext{c}{Aperture radius 8$\arcsec$}
\tablenotetext{d}{Aperture radius 15$\arcsec$}
\tablenotetext{e}{Aperture radius 8$\arcsec$}
\tablenotetext{f}{Aperture radius 20$\arcsec$}
\tablenotetext{g}{Aperture radius 35$\arcsec$}
\tablenotetext{h}{Aperture radius 5$\arcsec$}
\tablenotetext{i}{Aperture radius 10$\arcsec$}
\tablenotetext{j}{Aperture radius 14$\arcsec$}
\end{deluxetable}

\setcounter{table}{3}

\begin{deluxetable}{lccccccc}
\scriptsize
\tablewidth{0pt}
\tablecaption{Source Magnitudes}
\tablehead{
\colhead{Name} &
\colhead{B} &
\colhead{V} &
\colhead{R} &
\colhead{I} &
\colhead{J} &
\colhead{H} &
\colhead{K$_n$}
}
\startdata
PKS 0036$-$216 & $21.20 \pm 0.20$ & $19.36 \pm 0.07$ & $18.33 \pm 0.05$ & $17.46 \pm 0.05$ & $16.20 \pm 0.07$ & $15.39 \pm 0.06$ & $14.46 \pm 0.06 $ \nl
PKS 0038$-$020 & $18.74 \pm 0.07$ & $18.44 \pm 0.06$ & $18.11 \pm 0.06$ & $18.01 \pm 0.08$ & $17.44 \pm 0.15$ & $16.82 \pm 0.13$ & $16.16 \pm 0.13 $ \nl
PKS 0048$-$071 & $20.74 \pm 0.18$ & $20.32 \pm 0.13$ & $19.84 \pm 0.10$ & $19.53 \pm 0.15$ & $18.73 \pm 0.20$ & $17.93 \pm 0.16$ & $17.54 \pm 0.39 $ \nl
PKS 0048$-$097 & $16.12 \pm 0.05$ & $15.75 \pm 0.05$ & $15.39 \pm 0.05$ & $14.92 \pm 0.05$ & $13.99 \pm 0.05$ & $13.24 \pm 0.05$ & $12.51 \pm 0.05 $ \nl
PKS 0048$-$427 & $18.62 \pm 0.06$ & $18.45 \pm 0.06$ & $18.15 \pm 0.05$ & $17.75 \pm 0.06$ & $17.21 \pm 0.12$ & $16.86 \pm 0.14$ & $16.19 \pm 0.13 $ \nl
PKS 0056$-$001 & $17.73 \pm 0.05$ & $17.55 \pm 0.05$ & $17.43 \pm 0.05$ &      \nodata     & $16.49 \pm 0.08$ & $15.66 \pm 0.07$ & $14.92 \pm 0.06 $ \nl
PKS 0131$-$001 & $23.34 \pm 0.70$ & $22.50 \pm 0.50$ & $20.78 \pm 0.14$ & $20.17 \pm 0.17$ & $17.95 \pm 0.13$ & $17.24 \pm 0.12$ & $16.78 \pm 0.26 $ \nl
PKS 0153$-$410 & $19.81 \pm 0.10$ & $18.15 \pm 0.06$ & $17.41 \pm 0.05$ & $16.69 \pm 0.05$ & $15.49 \pm 0.06$ & $14.81 \pm 0.06$ & $14.14 \pm 0.06 $ \nl
PKS 0202$-$172 & $17.46 \pm 0.05$ & $17.33 \pm 0.05$ & $17.11 \pm 0.05$ & $16.74 \pm 0.05$ & $16.27 \pm 0.07$ & $15.67 \pm 0.07$ & $15.35 \pm 0.08 $ \nl
PKS 0213$-$026 & $21.33 \pm 0.20$ & $20.82 \pm 0.14$ & $19.99 \pm 0.08$ & $18.87 \pm 0.07$ & $17.48 \pm 0.15$ & $16.40 \pm 0.10$ & $15.17 \pm 0.07 $ \nl
PKS 0216$+$011 & $20.33 \pm 0.13$ & $20.36 \pm 0.13$ & $19.54 \pm 0.08$ & $19.02 \pm 0.09$ & $18.14 \pm 0.19$ & $17.55 \pm 0.12$ & $17.38 \pm 0.34 $ \nl
PKS 0220$-$349 & $21.73 \pm 0.38$ & $21.32 \pm 0.27$ & $20.83 \pm 0.20$ & $19.88 \pm 0.18$ & $18.72 \pm 0.20$ & $17.77 \pm 0.14$ & $16.82 \pm 0.22 $ \nl
PKS 0221$+$067 & $19.85 \pm 0.09$ & $18.97 \pm 0.06$ & $18.24 \pm 0.05$ & $17.36 \pm 0.05$ & $16.29 \pm 0.06$ & $15.26 \pm 0.06$ & $14.47 \pm 0.07 $ \nl
PKS 0226$-$038 & $17.56 \pm 0.05$ & $17.38 \pm 0.05$ & $17.17 \pm 0.05$ & $16.79 \pm 0.05$ & $16.40 \pm 0.07$ & $15.86 \pm 0.06$ & $15.13 \pm 0.06 $ \nl
PKS 0229$-$398 & $19.74 \pm 0.07$ & $20.31 \pm 0.10$ & $19.90 \pm 0.08$ & $19.46 \pm 0.10$ & $18.86 \pm 0.18$ & $17.55 \pm 0.12$ & $17.51 \pm 0.38 $ \nl
PKS 0232$-$042 & $15.73 \pm 0.05$ & $16.27 \pm 0.05$ & $15.93 \pm 0.05$ & $15.63 \pm 0.05$ & $15.22 \pm 0.05$ & $14.38 \pm 0.05$ & $14.28 \pm 0.05 $ \nl
PKS 0237$+$040 & $18.33 \pm 0.05$ & $17.88 \pm 0.05$ & $17.53 \pm 0.05$ & $17.14 \pm 0.05$ & $16.49 \pm 0.07$ & $15.97 \pm 0.07$ & $15.15 \pm 0.06 $ \nl
PKS 0238$-$084 & $12.03 \pm 0.05$ & $11.04 \pm 0.05$ & $10.44 \pm 0.05$ & $ 9.77 \pm 0.05$ & $ 8.70 \pm 0.05$ & $ 7.92 \pm 0.05$ & $ 7.68 \pm 0.05 $ \nl
PKS 0240$-$060 & $19.07 \pm 0.07$ & $18.87 \pm 0.07$ & $18.61 \pm 0.06$ & $18.07 \pm 0.07$ & $17.50 \pm 0.12$ & $17.33 \pm 0.16$ & $16.35 \pm 0.24 $ \nl
PKS 0240$-$217 & $19.20 \pm 0.09$ & $17.93 \pm 0.06$ & $17.07 \pm 0.05$ & $16.34 \pm 0.05$ &      \nodata     & $14.23 \pm 0.05$ & $13.63 \pm 0.06 $ \nl
PKS 0256$+$075 & $21.32 \pm 0.18$ & $20.68 \pm 0.13$ & $19.92 \pm 0.08$ & $18.92 \pm 0.08$ &      \nodata     &      \nodata     &      \nodata      \nl
PKS 0301$-$243 & $16.51 \pm 0.05$ & $16.10 \pm 0.05$ & $15.69 \pm 0.05$ & $15.22 \pm 0.05$ &      \nodata     &      \nodata     &      \nodata      \nl
PKS 0316$-$444 & $16.07 \pm 0.05$ & $14.76 \pm 0.05$ & $14.12 \pm 0.05$ & $13.51 \pm 0.05$ &      \nodata     &      \nodata     &      \nodata      \nl
PKS 0537$-$441 & $17.93 \pm 0.05$ & $17.34 \pm 0.05$ & $16.92 \pm 0.05$ & $16.35 \pm 0.05$ & $14.76 \pm 0.05$ & $13.85 \pm 0.05$ & $13.02 \pm 0.05 $ \nl
PKS 0829$+$046 & $16.28 \pm 0.05$ & $15.61 \pm 0.05$ & $15.11 \pm 0.05$ & $14.49 \pm 0.05$ & $13.45 \pm 0.05$ & $12.52 \pm 0.05$ & $11.80 \pm 0.05 $ \nl
PKS 0912$+$029 & $18.75 \pm 0.05$ & $18.59 \pm 0.06$ & $18.29 \pm 0.05$ & $17.87 \pm 0.06$ & $16.97 \pm 0.06$ & $16.01 \pm 0.05$ & $15.29 \pm 0.06 $ \nl
PKS 0922$+$005 &      \nodata     &      \nodata     & $17.09 \pm 0.05$ & $16.58 \pm 0.05$ &      \nodata     &      \nodata     &      \nodata      \nl
PKS 1016$-$311 & $18.32 \pm 0.06$ & $18.01 \pm 0.05$ & $17.82 \pm 0.05$ & $17.51 \pm 0.06$ & $16.70 \pm 0.08$ & $16.42 \pm 0.09$ & $15.59 \pm 0.14 $ \nl
PKS 1020$-$103 & $17.10 \pm 0.05$ & $16.70 \pm 0.05$ & $16.30 \pm 0.05$ & $15.62 \pm 0.05$ & $14.85 \pm 0.05$ & $14.11 \pm 0.05$ & $13.27 \pm 0.05 $ \nl
PKS 1021$-$006 & $18.38 \pm 0.06$ & $18.25 \pm 0.05$ & $18.03 \pm 0.05$ & $17.61 \pm 0.06$ & $17.12 \pm 0.10$ &      \nodata     &      \nodata      \nl
PKS 1032$-$199 & $18.92 \pm 0.06$ & $18.83 \pm 0.06$ & $18.60 \pm 0.06$ & $18.16 \pm 0.08$ & $17.35 \pm 0.12$ & $16.80 \pm 0.11$ & $16.04 \pm 0.10 $ \nl
PKS 1034$-$293 & $18.83 \pm 0.06$ & $18.26 \pm 0.05$ & $17.71 \pm 0.05$ & $17.01 \pm 0.05$ & $15.76 \pm 0.06$ & $14.83 \pm 0.05$ & $13.70 \pm 0.06 $ \nl
PKS 1036$-$154 & $20.54 \pm 0.11$ & $20.07 \pm 0.10$ & $19.43 \pm 0.08$ & $18.62 \pm 0.08$ & $17.23 \pm 0.11$ & $16.32 \pm 0.07$ & $15.63 \pm 0.15 $ \nl
PKS 1038$+$064 & $17.10 \pm 0.05$ & $16.92 \pm 0.05$ & $16.63 \pm 0.05$ & $16.43 \pm 0.05$ &      \nodata     &      \nodata     &      \nodata      \nl
PKS 1042$+$071 & $20.38 \pm 0.11$ & $19.87 \pm 0.11$ & $19.12 \pm 0.08$ & $18.27 \pm 0.08$ &      \nodata     &      \nodata     &      \nodata      \nl
PKS 1045$-$188 & $18.55 \pm 0.05$ & $18.26 \pm 0.05$ & $17.76 \pm 0.05$ & $17.08 \pm 0.05$ & $16.30 \pm 0.07$ & $15.30 \pm 0.06$ & $14.59 \pm 0.05 $ \nl
PKS 1048$-$313 & $20.19 \pm 0.10$ & $20.19 \pm 0.13$ & $19.70 \pm 0.11$ & $19.21 \pm 0.16$ &      \nodata     &      \nodata     &      \nodata      \nl
PKS 1055$-$243 & $19.13 \pm 0.07$ & $18.71 \pm 0.06$ & $18.40 \pm 0.06$ & $17.94 \pm 0.07$ &      \nodata     &      \nodata     &      \nodata      \nl
PKS 1055$+$018 & $18.17 \pm 0.05$ & $17.63 \pm 0.05$ & $17.15 \pm 0.05$ & $16.47 \pm 0.05$ &      \nodata     &      \nodata     &      \nodata      \nl
PKS 1101$-$325 & $15.54 \pm 0.05$ & $15.54 \pm 0.05$ & $15.31 \pm 0.05$ & $14.97 \pm 0.05$ & $14.55 \pm 0.05$ & $13.93 \pm 0.05$ & $13.16 \pm 0.05 $ \nl
PKS 1107$-$187 & $22.44 \pm 0.35$ & $21.10 \pm 0.14$ & $19.30 \pm 0.06$ & $18.74 \pm 0.07$ & $17.46 \pm 0.07$ & $17.21 \pm 0.07$ & $15.95 \pm 0.08 $ \nl
PKS 1110$-$217 & $24.41 \pm 1.08$ & $23.01 \pm 0.42$ & $21.57 \pm 0.13$ & $21.10 \pm 0.22$ & $19.06 \pm 0.19$ & $17.82 \pm 0.11$ & $16.74 \pm 0.15 $ \nl
PKS 1133$-$172 & $22.32 \pm 0.35$ & $21.54 \pm 0.19$ & $20.52 \pm 0.11$ & $19.41 \pm 0.09$ & $18.37 \pm 0.22$ & $17.69 \pm 0.19$ & $16.63 \pm 0.27 $ \nl
PKS 1136$-$135 & $16.65 \pm 0.05$ & $16.52 \pm 0.05$ & $16.39 \pm 0.05$ & $16.06 \pm 0.05$ & $15.54 \pm 0.05$ & $14.90 \pm 0.05$ & $14.40 \pm 0.05 $ \nl
PKS 1156$-$094 & $21.95 \pm 0.33$ & $21.43 \pm 0.22$ & $20.47 \pm 0.11$ & $20.22 \pm 0.17$ & $18.16 \pm 0.22$ & $17.51 \pm 0.20$ & $16.87 \pm 0.20 $ \nl
PKS 1244$-$255 & $17.42 \pm 0.05$ & $16.99 \pm 0.05$ & $16.63 \pm 0.05$ & $16.11 \pm 0.05$ & $15.43 \pm 0.05$ & $14.66 \pm 0.05$ & $13.85 \pm 0.06 $ \nl
PKS 1256$-$229 & $18.43 \pm 0.06$ & $17.68 \pm 0.05$ & $17.11 \pm 0.05$ & $16.39 \pm 0.05$ & $15.16 \pm 0.05$ & $14.21 \pm 0.05$ & $13.44 \pm 0.05 $ \nl
PKS 1313$-$333 & $18.17 \pm 0.06$ & $17.63 \pm 0.05$ & $17.16 \pm 0.05$ & $16.50 \pm 0.05$ & $15.46 \pm 0.05$ & $14.36 \pm 0.05$ & $13.69 \pm 0.06 $ \nl
PKS 1330$+$022 & $19.02 \pm 0.06$ & $17.89 \pm 0.05$ & $17.24 \pm 0.05$ & $16.47 \pm 0.05$ & $15.23 \pm 0.05$ & $14.43 \pm 0.05$ & $13.71 \pm 0.06 $ \nl
PKS 1353$-$341 & $19.20 \pm 0.06$ & $17.46 \pm 0.05$ & $16.74 \pm 0.05$ & $15.94 \pm 0.05$ & $15.20 \pm 0.06$ & $14.38 \pm 0.05$ & $13.62 \pm 0.06 $ \nl
PKS 1404$-$342 & $17.37 \pm 0.05$ & $16.78 \pm 0.05$ & $16.56 \pm 0.05$ & $16.35 \pm 0.05$ & $15.67 \pm 0.06$ & $15.13 \pm 0.06$ & $14.57 \pm 0.08 $ \nl
PKS 1411$+$094 & $18.64 \pm 0.09$ & $17.56 \pm 0.06$ & $16.77 \pm 0.05$ & $16.15 \pm 0.06$ & $15.10 \pm 0.07$ & $14.36 \pm 0.06$ & $13.78 \pm 0.06 $ \nl
PKS 1430$-$155 & $ > 22.70      $ & $23.24 \pm 0.48$ & $22.91 \pm 0.48$ & $20.73 \pm 0.16$ & $19.07 \pm 0.27$ & $17.97 \pm 0.13$ & $17.50 \pm 0.37 $ \nl
PKS 1430$-$178 & $19.14 \pm 0.07$ & $18.92 \pm 0.06$ & $18.55 \pm 0.06$ & $18.14 \pm 0.08$ & $17.75 \pm 0.24$ & $16.78 \pm 0.17$ & $15.69 \pm 0.23 $ \nl
PKS 1435$-$218 & $19.20 \pm 0.07$ & $18.54 \pm 0.06$ & $18.04 \pm 0.05$ & $17.75 \pm 0.06$ & $17.64 \pm 0.22$ & $17.18 \pm 0.23$ & $16.77 \pm 0.53 $ \nl
PKS 1437$-$153 & $19.94 \pm 0.11$ & $19.01 \pm 0.07$ & $18.47 \pm 0.06$ & $17.68 \pm 0.06$ & $16.29 \pm 0.08$ & $15.11 \pm 0.06$ & $14.24 \pm 0.08 $ \nl
PKS 1438$-$347 & $17.80 \pm 0.05$ & $17.31 \pm 0.05$ & $16.98 \pm 0.05$ & $16.72 \pm 0.05$ & $16.36 \pm 0.09$ & $15.86 \pm 0.09$ & $15.24 \pm 0.16 $ \nl
PKS 1450$-$338 & $22.52 \pm 0.37$ & $20.40 \pm 0.10$ & $19.39 \pm 0.07$ & $18.69 \pm 0.08$ & $16.99 \pm 0.08$ & $16.03 \pm 0.07$ & $15.23 \pm 0.09 $ \nl
PKS 1454$-$060 & $18.36 \pm 0.06$ & $17.84 \pm 0.05$ & $17.45 \pm 0.05$ & $17.08 \pm 0.05$ & $16.56 \pm 0.10$ & $15.86 \pm 0.09$ & $15.16 \pm 0.15 $ \nl
PKS 1456$+$044 &      \nodata     &      \nodata     & $18.83 \pm 0.05$ & $18.12 \pm 0.05$ & $16.17 \pm 0.05$ & $15.71 \pm 0.05$ & $14.30 \pm 0.05 $ \nl
PKS 1504$-$166 & $20.28 \pm 0.12$ & $19.75 \pm 0.11$ & $19.35 \pm 0.09$ & $18.46 \pm 0.10$ & $16.33 \pm 0.08$ & $15.59 \pm 0.07$ & $14.01 \pm 0.07 $ \nl
PKS 1508$-$055 & $17.37 \pm 0.05$ & $16.94 \pm 0.05$ & $16.66 \pm 0.05$ & $16.28 \pm 0.05$ & $15.56 \pm 0.06$ & $14.87 \pm 0.06$ & $14.29 \pm 0.08 $ \nl
PKS 1509$+$022 & $19.75 \pm 0.12$ & $18.27 \pm 0.06$ &      \nodata     &      \nodata     & $16.04 \pm 0.07$ & $15.12 \pm 0.06$ & $14.31 \pm 0.08 $ \nl
PKS 1510$-$089 & $17.26 \pm 0.05$ & $17.12 \pm 0.05$ & $16.72 \pm 0.05$ & $16.17 \pm 0.05$ & $15.24 \pm 0.05$ & $14.09 \pm 0.05$ & $13.27 \pm 0.05 $ \nl
PKS 1511$-$100 & $18.18 \pm 0.05$ & $18.00 \pm 0.05$ & $17.50 \pm 0.05$ & $17.07 \pm 0.05$ & $16.46 \pm 0.07$ & $15.60 \pm 0.06$ & $14.70 \pm 0.08 $ \nl
PKS 1511$-$210 & $22.01 \pm 0.25$ & $21.45 \pm 0.19$ & $20.65 \pm 0.11$ & $20.30 \pm 0.16$ & $18.62 \pm 0.32$ & $17.81 \pm 0.25$ & $16.79 \pm 0.36 $ \nl
PKS 1514$-$241 & $15.18 \pm 0.05$ & $14.50 \pm 0.05$ & $13.98 \pm 0.05$ & $13.31 \pm 0.05$ & $11.95 \pm 0.05$ & $11.10 \pm 0.05$ & $10.50 \pm 0.05 $ \nl
PKS 1518$+$045 & $16.11 \pm 0.05$ & $14.84 \pm 0.05$ & $14.20 \pm 0.05$ & $13.45 \pm 0.05$ & $12.06 \pm 0.05$ & $11.31 \pm 0.05$ & $11.02 \pm 0.05 $ \nl
PKS 1519$-$273 & $18.21 \pm 0.06$ & $17.74 \pm 0.05$ & $17.18 \pm 0.05$ & $16.55 \pm 0.05$ & $15.65 \pm 0.06$ & $14.84 \pm 0.06$ & $14.08 \pm 0.07 $ \nl
PKS 1532$+$016 & $19.12 \pm 0.08$ & $18.92 \pm 0.07$ & $18.44 \pm 0.06$ & $17.98 \pm 0.08$ & $17.40 \pm 0.18$ & $16.22 \pm 0.11$ & $15.48 \pm 0.19 $ \nl
PKS 1535$+$004 & $     > 21.70  $ &      \nodata     & $    > 22.30   $ & $   > 21.50    $ &      \nodata     & $    > 19.73   $ &      \nodata      \nl
PKS 1542$+$042 & $18.75 \pm 0.06$ & $18.54 \pm 0.06$ & $18.21 \pm 0.06$ & $17.61 \pm 0.06$ & $17.38 \pm 0.12$ & $16.46 \pm 0.09$ & $15.82 \pm 0.17 $ \nl
PKS 1546$+$027 & $17.43 \pm 0.05$ & $17.14 \pm 0.05$ & $16.75 \pm 0.05$ & $16.16 \pm 0.05$ & $14.97 \pm 0.05$ & $13.88 \pm 0.05$ & $13.02 \pm 0.05 $ \nl
PKS 1548$+$056 & $19.37 \pm 0.06$ & $18.65 \pm 0.06$ & $18.13 \pm 0.06$ & $17.42 \pm 0.06$ & $16.30 \pm 0.07$ & $15.13 \pm 0.06$ & $14.30 \pm 0.07 $ \nl
PKS 1550$-$269 & $20.24 \pm 0.14$ & $19.67 \pm 0.09$ & $19.18 \pm 0.08$ & $18.58 \pm 0.10$ & $17.63 \pm 0.15$ & $17.01 \pm 0.13$ & $16.07 \pm 0.11 $ \nl
PKS 1555$-$140 & $18.28 \pm 0.05$ & $16.93 \pm 0.05$ & $16.28 \pm 0.05$ & $15.47 \pm 0.05$ & $14.25 \pm 0.05$ & $13.43 \pm 0.05$ & $13.08 \pm 0.05 $ \nl
PKS 1555$+$001 & $20.34 \pm 0.10$ & $19.95 \pm 0.08$ & $20.00 \pm 0.07$ & $18.72 \pm 0.08$ & $17.73 \pm 0.16$ & $16.90 \pm 0.12$ & $16.24 \pm 0.24 $ \nl
PKS 1556$-$245 & $18.89 \pm 0.06$ & $19.02 \pm 0.06$ & $18.82 \pm 0.07$ & $18.30 \pm 0.08$ & $17.59 \pm 0.14$ & $16.97 \pm 0.13$ & $16.58 \pm 0.31 $ \nl
PKS 1601$-$222 &      \nodata     &      \nodata     & $    > 22.90   $ & $    > 22.40   $ &      \nodata     & $    > 20.02   $ & $    > 19.49    $ \nl
PKS 1602$-$001 & $17.66 \pm 0.05$ & $17.57 \pm 0.05$ & $17.10 \pm 0.05$ & $16.66 \pm 0.05$ & $16.28 \pm 0.07$ & $15.41 \pm 0.06$ & $15.19 \pm 0.08 $ \nl
PKS 1614$+$051 & $20.38 \pm 0.11$ & $19.38 \pm 0.07$ & $19.31 \pm 0.08$ & $18.93 \pm 0.10$ & $18.04 \pm 0.24$ & $17.83 \pm 0.30$ & $17.31 \pm 0.32 $ \nl
PKS 1615$+$029 & $18.01 \pm 0.05$ & $17.75 \pm 0.05$ & $17.31 \pm 0.05$ & $17.00 \pm 0.05$ & $16.66 \pm 0.05$ & $16.04 \pm 0.05$ & $15.89 \pm 0.08 $ \nl
PKS 1616$+$063 & $19.45 \pm 0.08$ & $18.97 \pm 0.06$ & $18.76 \pm 0.06$ & $18.27 \pm 0.08$ & $17.87 \pm 0.21$ & $17.54 \pm 0.24$ & $16.52 \pm 0.17 $ \nl
PKS 1635$-$035 & $21.58 \pm 0.32$ & $21.00 \pm 0.20$ & $20.75 \pm 0.18$ & $20.73 \pm 0.33$ & $19.11 \pm 0.18$ & $18.84 \pm 0.22$ & $17.48 \pm 0.32 $ \nl
PKS 1648$+$015 & $21.87 \pm 0.31$ & $21.14 \pm 0.18$ & $20.41 \pm 0.11$ & $19.50 \pm 0.10$ & $17.84 \pm 0.08$ & $16.85 \pm 0.06$ & $16.56 \pm 0.13 $ \nl
PKS 1649$-$062 &      \nodata     &      \nodata     & $    > 23.20   $ & $    > 23.30   $ & $    > 18.88   $ & $    > 17.77   $ & $    > 17.03    $ \nl
PKS 1654$-$020 & $23.51 \pm 0.68$ & $23.90 \pm 0.87$ & $23.05 \pm 0.52$ & $23.02 \pm 0.88$ & $    > 20.00   $ & $19.92 \pm 0.64$ & $18.25 \pm 0.56 $ \nl
PKS 1655$+$077 & $20.18 \pm 0.09$ & $19.88 \pm 0.07$ & $19.36 \pm 0.07$ & $18.42 \pm 0.07$ & $17.50 \pm 0.13$ & $16.59 \pm 0.10$ & $15.38 \pm 0.12 $ \nl
PKS 1656$+$053 & $17.33 \pm 0.05$ & $16.86 \pm 0.05$ & $16.64 \pm 0.05$ & $16.10 \pm 0.05$ & $15.54 \pm 0.06$ & $15.04 \pm 0.06$ & $14.39 \pm 0.09 $ \nl
PKS 1705$+$018 & $19.07 \pm 0.07$ & $18.66 \pm 0.06$ & $18.37 \pm 0.06$ & $17.86 \pm 0.07$ & $17.14 \pm 0.10$ & $16.44 \pm 0.09$ & $15.92 \pm 0.10 $ \nl
PKS 1706$+$006 & $22.15 \pm 0.29$ & $20.59 \pm 0.12$ & $19.95 \pm 0.10$ & $18.95 \pm 0.09$ & $17.52 \pm 0.16$ & $16.38 \pm 0.10$ & $15.63 \pm 0.17 $ \nl
PKS 1725$+$044 & $17.92 \pm 0.05$ & $17.46 \pm 0.05$ & $16.97 \pm 0.05$ & $16.37 \pm 0.05$ & $15.76 \pm 0.05$ & $14.91 \pm 0.05$ & $14.13 \pm 0.05 $ \nl
PKS 1732$+$094 & $    > 23.50   $ & $21.15 \pm 0.11$ & $19.74 \pm 0.06$ & $19.33 \pm 0.07$ & $18.40 \pm 0.20$ & $17.44 \pm 0.14$ & $16.37 \pm 0.20 $ \nl
PKS 1933$-$400 & $18.46 \pm 0.06$ & $18.03 \pm 0.05$ & $17.72 \pm 0.05$ & $17.43 \pm 0.06$ & $16.26 \pm 0.06$ & $15.79 \pm 0.07$ & $15.00 \pm 0.09 $ \nl
PKS 1953$-$325 & $19.60 \pm 0.09$ & $18.91 \pm 0.06$ & $18.52 \pm 0.06$ & $18.00 \pm 0.07$ & $17.32 \pm 0.16$ & $16.78 \pm 0.16$ & $15.92 \pm 0.26 $ \nl
PKS 1954$-$388 & $17.99 \pm 0.05$ & $17.55 \pm 0.05$ & $17.28 \pm 0.05$ & $16.89 \pm 0.05$ & $16.27 \pm 0.08$ & $15.28 \pm 0.06$ & $14.39 \pm 0.08 $ \nl
PKS 1958$-$179 & $19.48 \pm 0.08$ & $18.89 \pm 0.06$ & $18.34 \pm 0.06$ & $17.54 \pm 0.06$ & $16.26 \pm 0.08$ & $14.97 \pm 0.06$ & $14.39 \pm 0.08 $ \nl
PKS 2000$-$330 & $18.90 \pm 0.06$ & $17.48 \pm 0.05$ & $16.92 \pm 0.05$ & $16.71 \pm 0.05$ & $16.10 \pm 0.06$ & $15.64 \pm 0.06$ & $15.10 \pm 0.08 $ \nl
PKS 2002$-$185 & $19.39 \pm 0.07$ & $18.82 \pm 0.06$ & $18.55 \pm 0.06$ & $18.05 \pm 0.07$ & $17.03 \pm 0.10$ & $16.87 \pm 0.10$ & $16.92 \pm 0.33 $ \nl
PKS 2004$-$447 & $19.52 \pm 0.09$ & $18.74 \pm 0.06$ & $18.18 \pm 0.06$ & $17.35 \pm 0.06$ & $16.48 \pm 0.07$ & $15.46 \pm 0.06$ & $14.70 \pm 0.07 $ \nl
PKS 2008$-$159 & $17.50 \pm 0.05$ & $17.07 \pm 0.05$ & $16.76 \pm 0.05$ & $16.33 \pm 0.05$ & $16.18 \pm 0.08$ & $15.58 \pm 0.07$ & $14.97 \pm 0.12 $ \nl
PKS 2021$-$330 & $19.50 \pm 0.08$ & $19.06 \pm 0.07$ & $18.70 \pm 0.06$ & $18.31 \pm 0.08$ & $17.70 \pm 0.12$ & $16.82 \pm 0.09$ & $16.49 \pm 0.18 $ \nl
PKS 2022$-$077 & $19.10 \pm 0.06$ & $18.55 \pm 0.05$ & $18.44 \pm 0.05$ & $18.50 \pm 0.08$ & $18.11 \pm 0.30$ & $16.61 \pm 0.14$ & $16.08 \pm 0.30 $ \nl
PKS 2037$-$253 & $19.69 \pm 0.09$ & $19.56 \pm 0.09$ & $19.21 \pm 0.08$ & $19.24 \pm 0.16$ & $18.52 \pm 0.34$ & $18.01 \pm 0.34$ & $17.40 \pm 0.66 $ \nl
PKS 2044$-$168 & $17.70 \pm 0.05$ & $17.55 \pm 0.05$ & $17.18 \pm 0.05$ & $17.05 \pm 0.05$ & $16.48 \pm 0.09$ & $15.99 \pm 0.09$ & $15.18 \pm 0.15 $ \nl
PKS 2047$+$098 &      \nodata     &      \nodata     & $    > 22.99   $ & $    > 22.19   $ &      \nodata     & $    > 19.92   $ &      \nodata      \nl
PKS 2053$-$044 & $18.86 \pm 0.06$ & $18.47 \pm 0.06$ & $18.18 \pm 0.06$ & $17.99 \pm 0.07$ & $17.81 \pm 0.20$ & $16.96 \pm 0.15$ & $16.13 \pm 0.25 $ \nl
PKS 2056$-$369 &      \nodata     & $    > 23.50   $ & $23.45 \pm 0.61$ & $22.22 \pm 0.44$ & $19.55 \pm 0.39$ & $19.07 \pm 0.39$ & $18.21 \pm 0.41 $ \nl
PKS 2058$-$135 & $15.88 \pm 0.05$ & $14.76 \pm 0.05$ & $14.15 \pm 0.05$ & $13.45 \pm 0.05$ & $12.35 \pm 0.05$ & $11.58 \pm 0.05$ & $11.39 \pm 0.05 $ \nl
PKS 2058$-$297 & $18.85 \pm 0.06$ & $18.68 \pm 0.06$ & $18.29 \pm 0.06$ & $18.09 \pm 0.07$ & $17.52 \pm 0.15$ & $17.00 \pm 0.15$ & $16.22 \pm 0.35 $ \nl
PKS 2059$+$034 & $18.09 \pm 0.05$ & $17.65 \pm 0.05$ & $17.40 \pm 0.05$ & $17.20 \pm 0.06$ & $16.34 \pm 0.07$ & $16.18 \pm 0.08$ & $15.63 \pm 0.15 $ \nl
PKS 2106$-$413 & $19.57 \pm 0.09$ & $18.85 \pm 0.06$ & $18.45 \pm 0.06$ & $17.76 \pm 0.06$ & $16.77 \pm 0.08$ & $15.74 \pm 0.06$ & $14.81 \pm 0.08 $ \nl
PKS 2120$+$099 & $19.65 \pm 0.10$ & $19.24 \pm 0.08$ & $18.96 \pm 0.08$ & $18.36 \pm 0.10$ & $17.98 \pm 0.23$ & $17.66 \pm 0.26$ & $16.52 \pm 0.34 $ \nl
PKS 2121$+$053 & $19.86 \pm 0.10$ & $19.22 \pm 0.07$ & $18.68 \pm 0.06$ & $17.93 \pm 0.07$ & $16.52 \pm 0.07$ & $15.49 \pm 0.06$ & $14.71 \pm 0.08 $ \nl
PKS 2126$-$158 & $17.97 \pm 0.05$ & $16.92 \pm 0.05$ & $16.63 \pm 0.05$ & $16.23 \pm 0.05$ & $15.45 \pm 0.05$ & $14.89 \pm 0.05$ & $14.30 \pm 0.05 $ \nl
PKS 2127$-$096 & $19.43 \pm 0.07$ & $18.65 \pm 0.06$ & $18.00 \pm 0.06$ & $17.39 \pm 0.06$ & $16.38 \pm 0.07$ & $15.41 \pm 0.06$ & $14.61 \pm 0.08 $ \nl
PKS 2128$-$123 & $15.28 \pm 0.05$ & $15.13 \pm 0.05$ & $15.00 \pm 0.05$ & $14.70 \pm 0.05$ & $14.58 \pm 0.05$ & $13.89 \pm 0.05$ & $13.22 \pm 0.05 $ \nl
PKS 2131$-$021 & $20.61 \pm 0.13$ & $20.16 \pm 0.15$ & $19.58 \pm 0.11$ & $19.11 \pm 0.17$ & $17.58 \pm 0.14$ & $16.71 \pm 0.11$ & $15.85 \pm 0.17 $ \nl
PKS 2134$+$004 & $17.04 \pm 0.05$ & $16.80 \pm 0.05$ & $16.56 \pm 0.05$ & $16.11 \pm 0.05$ & $15.67 \pm 0.06$ & $15.18 \pm 0.06$ & $14.73 \pm 0.08 $ \nl
PKS 2135$-$248 & $19.64 \pm 0.10$ & $19.23 \pm 0.08$ & $18.96 \pm 0.08$ & $18.41 \pm 0.10$ & $17.50 \pm 0.13$ & $16.76 \pm 0.11$ & $15.89 \pm 0.18 $ \nl
PKS 2140$-$048 & $17.22 \pm 0.05$ & $17.18 \pm 0.05$ & $16.87 \pm 0.05$ & $16.44 \pm 0.05$ & $15.59 \pm 0.05$ & $14.76 \pm 0.05$ & $13.95 \pm 0.06 $ \nl
PKS 2143$-$156 & $18.02 \pm 0.05$ & $17.69 \pm 0.05$ & $17.55 \pm 0.05$ & $17.09 \pm 0.06$ & $16.61 \pm 0.08$ & $15.91 \pm 0.07$ & $15.24 \pm 0.11 $ \nl
PKS 2144$+$092 & $18.78 \pm 0.07$ & $18.36 \pm 0.06$ & $18.01 \pm 0.06$ & $17.68 \pm 0.07$ & $16.37 \pm 0.07$ & $15.69 \pm 0.07$ & $14.82 \pm 0.09 $ \nl
PKS 2145$-$176 & $19.43 \pm 0.09$ & $19.11 \pm 0.08$ & $18.85 \pm 0.07$ & $18.23 \pm 0.09$ & $17.74 \pm 0.19$ & $17.25 \pm 0.19$ & $16.41 \pm 0.32 $ \nl
PKS 2145$+$067 & $16.15 \pm 0.05$ & $15.80 \pm 0.05$ & $15.56 \pm 0.05$ & $15.29 \pm 0.05$ & $14.54 \pm 0.05$ & $14.17 \pm 0.05$ & $13.48 \pm 0.06 $ \nl
PKS 2149$-$307 & $18.00 \pm 0.05$ & $17.75 \pm 0.05$ & $17.31 \pm 0.05$ & $17.20 \pm 0.06$ & $16.60 \pm 0.08$ & $16.02 \pm 0.08$ & $15.24 \pm 0.13 $ \nl
PKS 2149$+$056 & $    > 23.50   $ & $22.05 \pm 0.23$ & $20.85 \pm 0.11$ & $19.54 \pm 0.10$ & $18.26 \pm 0.24$ & $17.42 \pm 0.18$ & $17.17 \pm 0.49 $ \nl
PKS 2149$+$069 & $18.79 \pm 0.07$ & $18.48 \pm 0.06$ & $17.99 \pm 0.06$ & $17.74 \pm 0.06$ & $16.90 \pm 0.10$ & $15.82 \pm 0.07$ & $15.76 \pm 0.19 $ \nl
PKS 2155$-$152 & $18.42 \pm 0.05$ & $17.90 \pm 0.05$ & $17.45 \pm 0.05$ & $16.88 \pm 0.05$ & $15.67 \pm 0.05$ & $14.72 \pm 0.05$ & $13.92 \pm 0.05 $ \nl
PKS 2200$-$238 & $18.45 \pm 0.06$ & $18.36 \pm 0.06$ & $18.10 \pm 0.06$ & $17.64 \pm 0.07$ & $17.27 \pm 0.11$ & $16.72 \pm 0.11$ & $15.82 \pm 0.09 $ \nl
PKS 2203$-$188 & $18.85 \pm 0.07$ & $18.52 \pm 0.06$ & $18.10 \pm 0.06$ & $17.46 \pm 0.06$ & $16.69 \pm 0.08$ & $15.93 \pm 0.07$ & $15.25 \pm 0.07 $ \nl
PKS 2206$-$237 & $17.41 \pm 0.06$ & $16.25 \pm 0.05$ & $15.70 \pm 0.05$ & $15.04 \pm 0.05$ & $13.94 \pm 0.05$ & $13.23 \pm 0.05$ & $12.87 \pm 0.06 $ \nl
PKS 2208$-$137 & $16.79 \pm 0.05$ & $16.70 \pm 0.05$ & $16.42 \pm 0.05$ & $16.10 \pm 0.05$ & $15.13 \pm 0.05$ & $14.34 \pm 0.05$ & $13.40 \pm 0.05 $ \nl
PKS 2210$-$257 & $18.74 \pm 0.07$ & $18.76 \pm 0.07$ & $18.53 \pm 0.06$ & $17.95 \pm 0.07$ & $17.34 \pm 0.12$ & $16.42 \pm 0.09$ & $15.65 \pm 0.08 $ \nl
PKS 2212$-$299 & $17.47 \pm 0.05$ & $17.36 \pm 0.05$ & $17.12 \pm 0.05$ & $16.81 \pm 0.05$ & $15.92 \pm 0.06$ & $15.19 \pm 0.06$ & $14.59 \pm 0.06 $ \nl
PKS 2215$+$020 & $21.84 \pm 0.31$ & $20.42 \pm 0.10$ & $20.14 \pm 0.12$ & $20.00 \pm 0.20$ & $19.20 \pm 0.50$ & $18.21 \pm 0.29$ & $19.34 \pm 1.78 $ \nl
PKS 2216$-$038 & $17.03 \pm 0.05$ & $16.49 \pm 0.05$ & $16.14 \pm 0.05$ & $15.80 \pm 0.05$ & $14.81 \pm 0.05$ & $14.42 \pm 0.05$ & $13.96 \pm 0.05 $ \nl
PKS 2223$-$052 & $18.59 \pm 0.06$ & $18.33 \pm 0.06$ & $17.89 \pm 0.05$ & $17.43 \pm 0.06$ & $16.45 \pm 0.07$ & $15.51 \pm 0.06$ & $14.69 \pm 0.06 $ \nl
PKS 2227$-$088 & $18.31 \pm 0.06$ & $18.15 \pm 0.06$ & $17.81 \pm 0.05$ & $17.39 \pm 0.06$ & $16.47 \pm 0.07$ & $15.70 \pm 0.06$ & $15.06 \pm 0.09 $ \nl
PKS 2227$-$399 & $17.94 \pm 0.05$ & $17.77 \pm 0.05$ & $17.19 \pm 0.05$ & $16.69 \pm 0.05$ & $15.87 \pm 0.06$ & $15.02 \pm 0.05$ & $14.11 \pm 0.06 $ \nl
PKS 2229$-$172 & $21.26 \pm 0.19$ & $21.14 \pm 0.18$ & $21.34 \pm 0.30$ & $19.83 \pm 0.17$ & $18.88 \pm 0.27$ & $17.71 \pm 0.21$ & $16.77 \pm 0.17 $ \nl
PKS 2233$-$148 & $19.24 \pm 0.06$ & $18.61 \pm 0.06$ & $18.06 \pm 0.05$ & $17.27 \pm 0.05$ & $16.38 \pm 0.06$ & $15.42 \pm 0.06$ & $14.57 \pm 0.05 $ \nl
PKS 2239$+$096 & $19.00 \pm 0.06$ & $18.68 \pm 0.06$ & $18.34 \pm 0.06$ & $17.91 \pm 0.06$ & $17.42 \pm 0.12$ & $16.73 \pm 0.10$ & $16.01 \pm 0.18 $ \nl
PKS 2240$-$260 & $17.95 \pm 0.05$ & $17.37 \pm 0.05$ & $16.95 \pm 0.05$ & $16.36 \pm 0.05$ & $15.15 \pm 0.05$ & $14.32 \pm 0.05$ & $13.52 \pm 0.05 $ \nl
PKS 2243$-$123 & $16.65 \pm 0.05$ & $16.50 \pm 0.05$ & $16.32 \pm 0.05$ & $16.04 \pm 0.05$ & $15.51 \pm 0.05$ & $15.00 \pm 0.05$ & $14.50 \pm 0.07 $ \nl
PKS 2245$-$328 & $19.08 \pm 0.07$ & $18.82 \pm 0.07$ & $18.51 \pm 0.06$ & $18.31 \pm 0.08$ & $17.61 \pm 0.13$ & $17.20 \pm 0.14$ & $16.08 \pm 0.19 $ \nl
PKS 2245$+$029 & $21.71 \pm 0.22$ & $21.00 \pm 0.17$ & $20.26 \pm 0.10$ & $19.48 \pm 0.11$ & $17.45 \pm 0.14$ & $16.63 \pm 0.12$ & $15.49 \pm 0.15 $ \nl
PKS 2252$-$090 & $22.03 \pm 0.24$ & $21.51 \pm 0.16$ & $20.49 \pm 0.12$ & $19.48 \pm 0.11$ & $18.98 \pm 0.49$ & $17.02 \pm 0.16$ & $16.48 \pm 0.33 $ \nl
PKS 2312$-$319 & $18.10 \pm 0.05$ & $17.86 \pm 0.05$ & $17.50 \pm 0.05$ & $17.09 \pm 0.05$ & $16.82 \pm 0.09$ & $16.04 \pm 0.08$ & $15.63 \pm 0.09 $ \nl
PKS 2313$-$438 & $19.89 \pm 0.11$ & $19.53 \pm 0.08$ & $19.10 \pm 0.07$ & $18.60 \pm 0.10$ & $17.97 \pm 0.19$ & $17.37 \pm 0.18$ & $16.72 \pm 0.17 $ \nl
PKS 2314$-$409 & $18.33 \pm 0.06$ & $18.08 \pm 0.05$ & $17.75 \pm 0.05$ & $17.36 \pm 0.06$ & $16.74 \pm 0.08$ & $16.15 \pm 0.08$ & $15.59 \pm 0.08 $ \nl
PKS 2329$-$415 & $19.40 \pm 0.08$ & $19.17 \pm 0.07$ & $18.89 \pm 0.07$ & $18.41 \pm 0.09$ & $17.58 \pm 0.14$ & $16.67 \pm 0.10$ & $15.67 \pm 0.08 $ \nl
PKS 2337$-$334 & $22.93 \pm 0.60$ & $21.89 \pm 0.27$ & $21.40 \pm 0.20$ & $20.51 \pm 0.14$ & $19.88 \pm 0.54$ & $18.42 \pm 0.32$ & $16.39 \pm 0.20 $ \nl
PKS 2344$-$192 & $23.52 \pm 0.68$ &      \nodata     & $21.68 \pm 0.25$ & $20.96 \pm 0.27$ & $18.71 \pm 0.13$ & $17.64 \pm 0.09$ & $17.25 \pm 0.20 $ \nl
PKS 2345$-$167 & $17.23 \pm 0.05$ & $16.79 \pm 0.05$ & $16.44 \pm 0.05$ & $15.88 \pm 0.05$ & $14.53 \pm 0.05$ & $13.57 \pm 0.05$ & $12.76 \pm 0.05 $ \nl
PKS 2351$-$154 & $18.97 \pm 0.07$ & $18.71 \pm 0.06$ & $18.46 \pm 0.06$ & $18.42 \pm 0.10$ & $17.88 \pm 0.18$ & $17.06 \pm 0.14$ & $16.37 \pm 0.26 $ \nl
PKS 2354$-$117 & $18.81 \pm 0.07$ & $18.03 \pm 0.05$ & $17.80 \pm 0.05$ & $17.44 \pm 0.06$ & $16.56 \pm 0.07$ & $15.96 \pm 0.07$ & $15.06 \pm 0.06 $ \nl
\tablevspace{2mm}
L 2110$-$4509  & $18.89 \pm 0.07$ & $18.64 \pm 0.07$ & $18.21 \pm 0.06$ & $17.78 \pm 0.08$ & $16.19 \pm 0.06$ & $15.37 \pm 0.06$ & $15.13 \pm 0.10 $ \nl
L 2111$-$4335  & $17.29 \pm 0.05$ & $16.78 \pm 0.05$ & $16.54 \pm 0.05$ & $15.99 \pm 0.05$ & $15.25 \pm 0.05$ & $14.55 \pm 0.05$ & $14.33 \pm 0.07 $ \nl
L 2111$-$4506  & $17.54 \pm 0.05$ & $17.37 \pm 0.05$ & $16.94 \pm 0.05$ & $16.73 \pm 0.05$ & $16.16 \pm 0.06$ & $15.35 \pm 0.06$ & $15.32 \pm 0.12 $ \nl
L 2113$-$4305  & $17.87 \pm 0.05$ & $17.62 \pm 0.05$ & $17.18 \pm 0.05$ & $16.92 \pm 0.06$ & $16.39 \pm 0.07$ & $15.96 \pm 0.07$ & $15.21 \pm 0.11 $ \nl
L 2113$-$4345  & $18.56 \pm 0.06$ & $18.22 \pm 0.06$ & $17.99 \pm 0.06$ & $17.62 \pm 0.07$ & $17.12 \pm 0.10$ & $16.86 \pm 0.12$ & $16.21 \pm 0.23 $ \nl
L 2113$-$4538  & $17.67 \pm 0.05$ & $17.49 \pm 0.05$ & $17.27 \pm 0.05$ & $16.89 \pm 0.06$ &      \nodata     &      \nodata     &      \nodata      \nl
L 2114$-$4335  & $18.22 \pm 0.06$ & $17.90 \pm 0.05$ & $17.52 \pm 0.05$ & $17.25 \pm 0.05$ &      \nodata     &      \nodata     &      \nodata      \nl
L 2114$-$4346  & $18.56 \pm 0.06$ & $18.42 \pm 0.06$ & $18.09 \pm 0.06$ & $17.79 \pm 0.06$ &      \nodata     &      \nodata     &      \nodata      \nl
L 2114$-$4501  & $18.67 \pm 0.06$ & $18.54 \pm 0.06$ & $18.49 \pm 0.06$ & $18.05 \pm 0.07$ &      \nodata     &      \nodata     &      \nodata      \nl
L 2116$-$4439  & $18.61 \pm 0.06$ & $18.20 \pm 0.06$ & $17.80 \pm 0.05$ & $17.40 \pm 0.06$ &      \nodata     &      \nodata     &      \nodata      \nl
L 2118$-$4702  & $19.21 \pm 0.08$ & $18.98 \pm 0.07$ & $18.64 \pm 0.06$ & $18.40 \pm 0.08$ &      \nodata     &      \nodata     & $16.67 \pm 0.30 $ \nl
L 2119$-$4415  & $18.02 \pm 0.05$ & $17.72 \pm 0.05$ & $17.43 \pm 0.05$ & $16.88 \pm 0.05$ & $16.07 \pm 0.06$ & $15.68 \pm 0.06$ & $15.03 \pm 0.09 $ \nl
\enddata
\end{deluxetable}